\documentclass[fleqn,usenatbib]{mnras}

\usepackage{newtxtext,newtxmath}

\usepackage[T1]{fontenc}
\usepackage{ae,aecompl}

\usepackage{graphicx}	
\usepackage{amsmath}	
\usepackage{footnote}
\usepackage[bottom, flushmargin]{footmisc}

\title[Characterizing HD~165246]{Characterization of the variability in the O+B eclipsing binary HD~165246\thanks{Based on observations obtained with the HERMES spectrograph on the Mercator telescope, which is supported by the Research Foundation - Flanders (FWO), Belgium, the Research Council of KU Leuven, Belgium, the Fonds National de la Recherche Scientifique (F.R.S.-FNRS), Belgium, the Royal Observatory of Belgium, the Observatoire de Gen{\'e}ve, Switzerland and the Th{\"u}ringer Landessternwarte Tautenburg, Germany.}}
\author[C. Johnston et al.]
{C.~Johnston$^{1,2}$\thanks{E-mail: colecampbell.johnston@kuleuven.be},
N.~Aimar$^{1,3}$,
M.~Abdul-Masih$^{4,1}$,
D.~M. Bowman$^{1}$,
T.~R. White$^{5,6}$,
\newauthor
C. Hawcroft$^{1}$,
H.~Sana$^{1}$,
S.~Sekaran$^{1}$,
K.~Dsilva$^{1}$,
A.~Tkachenko$^{1}$,
C.~Aerts$^{1,2,7}$
\\
$^{1}$Institute of Astronomy, KU Leuven, Celestijnenlaan 200D, 3001
  Leuven, Belgium\\
$^{2}$Department of Astrophysics, IMAPP, Radboud University Nijmegen,
P. O. Box 9010, 6500 GL Nijmegen, the Netherlands\\
$^{3}$Observatoire de Paris, Universit{\'e} PSL, 5 Place Jules Janssen, 92195 Meudon, France\\
$^{4}$European Southern Observatory, Alonso de C{\' o}rdova 3107, Vitacura, Casilla 19001, Santiago de Chile, Chile\\
$^{5}$Sydney Institute for Astronomy (SIfA), School of Physics, University of Sydney, NSW 2006, Australia\\
$^{6}$Stellar Astrophysics Centre, Department of Physics and Astronomy, Aarhus University, Ny Munkegade 120, DK-8000 Aarhus C, Denmark\\
$^{7}$Max Planck Institute for Astronomy, Koenigstuhl 17, 69117 Heidelberg, Germany \\
}

\date{Accepted XXX. Received YYY; in original form ZZZ}

\pubyear{2020}

\begin{document}
\label{firstpage}
\pagerange{\pageref{firstpage}--\pageref{lastpage}}
\maketitle

\begin{abstract}
O-stars are known to experience a wide range of variability mechanisms originating
at both their surface and their near-core regions. Characterization and understanding
of this variability and its potential causes are integral for evolutionary calculations.
We use a new extensive high-resolution spectroscopic data set to characterize the
variability observed in both the spectroscopic and space-based photometric observations
of the O+B eclipsing binary HD~165246. We present an updated atmospheric and binary
solution for the primary component, involving a high level of microturbulence
($13_{-1.3}^{+1.0}\,$km\,s$^{-1}$) and a mass of $M_1=23.7_{-1.4}^{+1.1}$~M$_{\odot}$,
placing it in a sparsely explored region of the Hertzsprung-Russell diagram. Furthermore,
we deduce a rotational frequency of $0.690\pm 0.003\,$d$^{-1}$ from the combined
photometric and line-profile variability, implying that the primary rotates at 40\% of
its critical Keplerian rotation rate. We discuss the potential explanations for the
overall variability observed in this massive binary, and discuss its evolutionary context.
\end{abstract}




\section{Introduction}
Despite the impact that massive stars ($M\mathrm{>9\,M_{\odot}}$) have on
the dynamical and chemical evolution of their environment and host galaxy,
the physics that determine their evolution are poorly calibrated in theoretical
stellar structure and evolutionary models \citep{Kippenhahn2012,Langer2012}. In
particular, these uncalibrated physics propagate into population synthesis
predictions, chemical evolutionary models, and gravitational-wave progenitor predictions.
As such, calibrating the input physics that govern massive star evolution
is an important goal of astrophysical research. However, due to the diversity
of mechanisms that may or may not be active in a given star, calibrating
evolutionary models of massive stars can be difficult.

Massive stars exhibit a multitude of phenomena, including
(in some cases extreme) wind-mass loss \citep[e.g.,][]{Vink2011,Sander2019}, rapid rotation
\citep[e.g.,][]{Maeder2009,Ekstrom2012,deMink2013,Abdulmasih2019}, magnetic fields
\citep[e.g.,][]{Wade2016,Buysschaert2017a}, as well as pulsations \citep[e.g.,][]{Aerts2010,
Handler2013,Bowman2020b}.
Additionally, massive O- and B-stars are observed to have a high binary or multiplicity fraction
\citep{Kiminki2012,Sana2012,Sana2013,Sana2014,Aldoretta2015}. Large scale spectroscopic
surveys have attempted to populate the upper main-sequence (MS) region of the Hertzsprung-Russell
diagram (HRD) in order to characterize the variability there, but face the challenge of disentangling
multiple sources of variability \citep{SimonDiaz2017,Godart2017,Burssens2020}.
All of these variability mechanisms manifest spectroscopically and photometrically,
with their observable effects being useful in constraining the physics responsible for them.

Stars with initial masses between approximately 8 and 25~M$_{\odot}$, and between 3 and
9~M$_{\odot}$ have been observed to pulsate in coherent pressure (p) and gravity (g) modes
\citep{Aerts2010} excited by the $\kappa$-mechanism \citep{Pamyatnykh1999}, with their dominant
restoring forces being the pressure force and the buoyancy force, respectively. Historically, these
pulsators were difficult to study from the ground given the periods involved, however, long time-base,
high duty-cycle space based missions have revealed some of these massive pulsators to be multi-periodic
hybrid pulsators \citep{deCat2004,Briquet2007,Daszynska2010,Szewczuk2017,Burssens2019}. In addition
to coherent pulsations excited via the $\kappa$-mechanism, OB stars are also observed to exhibit
internal gravity waves \citep[IGWs;][]{Aerts2015b, Aerts2017,Bowman2019a,Bowman2019b}, which are
stochastically excited travelling buoyancy waves generated at the interface of the convective core
and radiative envelope. Irrespective of the excitation mechanism, these waves cause both spectroscopic
and photometric variability in observations \citep[see e.g.][]{Bowman2020a}. Additionally, the
photometric signal associated with IGWs has been independently reproduced via both 2D
\citep{Ratnasingam2020,Horst2020} and 3D hydrodynamical simulations \citep{Edelmann2019}.

Whenever active, any of these mechanisms can substantially impact the evolution
of a star and its resulting end product. To this end, characterization of the signals
present in observations of massive stars is required before they can unambiguously be
attributed to some given mechanism(s). As such, massive stars in eclipsing binaries
represent some of the best opportunities to achieve this goal, as they provide
model-independent estimates of fundamental stellar parameters such as mass and radius.
However, the currently-known sample of massive eclipsing binaries with data sets that
allow for such precise parameter determination is limited \citep{Torres2010,Bonanos2011,
Koumpia2012,Lohr2018,Mahy2020a,Mahy2020b}. Even more so, there are fewer systems for
which there exists extensive photometric and spectroscopic data sets to characterize the
variety of variability.

In this paper, we investigate and characterize the intrinsic variability in the
massive detached eclipsing O+B binary HD~165246. To do this, we utilise an
extensive spectroscopic data set combined with high-precision space photometry. In
Section~\ref{section:photometric_analysis} we determine an updated binary solution,
and fundamental parameters for the system. Furthermore, in Section~\ref{section:ipw_analysis}
we investigate the photometric variability present in the residual light curve after
subtraction of the binary model. In Section~\ref{section:spectroscopic_analysis},
we introduce {and analyse} the new spectroscopic data set. Finally, in
Section~\ref{section:discussion}, we discuss the potential mechanisms behind the
observed variability and briefly discuss the evolutionary context of the system,
and present our conclusions in Section~\ref{section:conclusions}.

\section{The target HD~165246}
\label{section:previous_studies}

HD~165246 ($V=7.6$, O8V) has been the focus of several dedicated studies and has
been included as part of a larger sample in some spectroscopic surveys of massive stars.
\citet{Mayer2013} made use of 13 FEROS spectra and 617 ASAS3 V-band observations to
derive an initial orbital solution for this system, improving on the linear ephemeris
deduced by \citet{Otero2007}. \citet{Mayer2013} determined a projected rotational
velocity of $v\sin i=242.6\pm2.7$~km~s$^{-1}$, without including micro- or macroturbulence.
In their analysis, \citet{Mayer2013} assumed a mass
for the primary of $M_1=21.5~M_{\odot}$, obtained by comparing the spectroscopic
parameters to theoretical models by \citet{Martins2005}. From this, the authors
determine a full spectroscopic orbital solution via disentangling, enforcing
a mass ratio of $q=0.175$ as a starting point. They obtained $T_{\rm eff,1}=33\,300\pm400$~K
and $T_{\rm eff,2}=15\,800\pm 700$~K. Furthermore, \citet{Mayer2013} claimed that while the
derived primary radius is typical for an O8V star, the solution for the secondary revealed
a smaller radius than predicted for its mass. From their solution, \citet{Mayer2013} determined
an age of $\tau=3.3\pm0.2$~Myr via comparison with rotating theoretical evolutionary tracks
\citep{Brott2011a}.

Using SAM/NACO measurements obtained with the VLTI, \citet{Sana2014} detected a close
bright ($\Delta H_{\rm mag}=2.36$) companion to the inner binary at a separation of
$\rho=30\pm16$~mas, and confirmed the presence of a bright companion ($\Delta H_{\rm mag}=3.36$)
at a separation of $\rho=1.93\pm0.04$~arcsec that was originally detected by \citet{Mason1998}.
Additionally, \citet{Sana2014} detected two fainter companions ($\Delta H_{\rm mag} > 5$)
at separations larger than 6.5 arcsec. Including the two distant fainter companions,
HD~165246 is a sextuple system with the 4.6-d O+B eclipsing binary. Without further
interferometric observations, additional components interior to the companion at 30 mas
cannot be ruled out.

After the failure of a second reaction wheel in 2013, the $Kepler$ satellite was re-purposed
as the $K2$ mission, which scanned the ecliptic in 90-d long pointings \citep{Howell2014}.
Owing to the densely crowded field in $K2$ Campaign 9, \citet{Johnston2017} extracted a
$\sim$30-d light curve using a custom aperture mask, remarking that the mildly saturated
pixels, systematics, and thruster firings rendered the remaining $\sim$60-d segment of the lightcurve
of too poor quality for science. Assuming a circular orbit, and adopting the mass ratio and
primary effective temperature reported by \citet{Mayer2013}, \citet{Johnston2017} determined
an updated binary model. Investigation of the residuals after subtraction of the optimised binary model
revealed multi-periodic variability on time scales between several hours and a few days.
Additionally, high-order harmonics of the orbital frequency as well as a harmonic series whose base
frequency is consistent with the rotational frequency of the primary were observed. However,
the orbital harmonics may have originated from the assumed circular orbit, which can be mitigated
with a larger spectroscopic data set. The authors also note the possibility of Doppler beaming being
present, but relegate investigation to future work.

From their binary model, \citet{Johnston2017} estimated 18~per~cent contaminating light in the
light curve, which is within the 7 to 43~per~cent third light contribution estimated from the
magnitude contrasts for the other members of the sextuple system \citep{Sana2014}. This estimate
depends on the primary effective temperature and mass ratio adopted from \citet{Mayer2013},
and as such is subject to change. Finally, \citet{Johnston2017} concluded that the variability
signal likely originates from the primary, but suggest that further follow-up is required for
unambiguous characterization.

\section{Photometric analysis}
\label{section:photometric_analysis}

The process of optimising the binary and atmospheric solutions is an iterative one, with
the binary model relying on information determined from an atmospheric solution and vice
versa. Hence, we start with an atmospheric solution, then build a binary model, and
repeat the atmospheric modelling with the updated information from the binary model.
This process is repeated until both solutions no longer change within the output
uncertainties. For clarity, we describe the binary and atmospheric processes separately.

\subsection{K2 Photometry}
\label{section:k2_photometry}

\begin{figure*}
    \centering
    \includegraphics[width=0.875\linewidth]{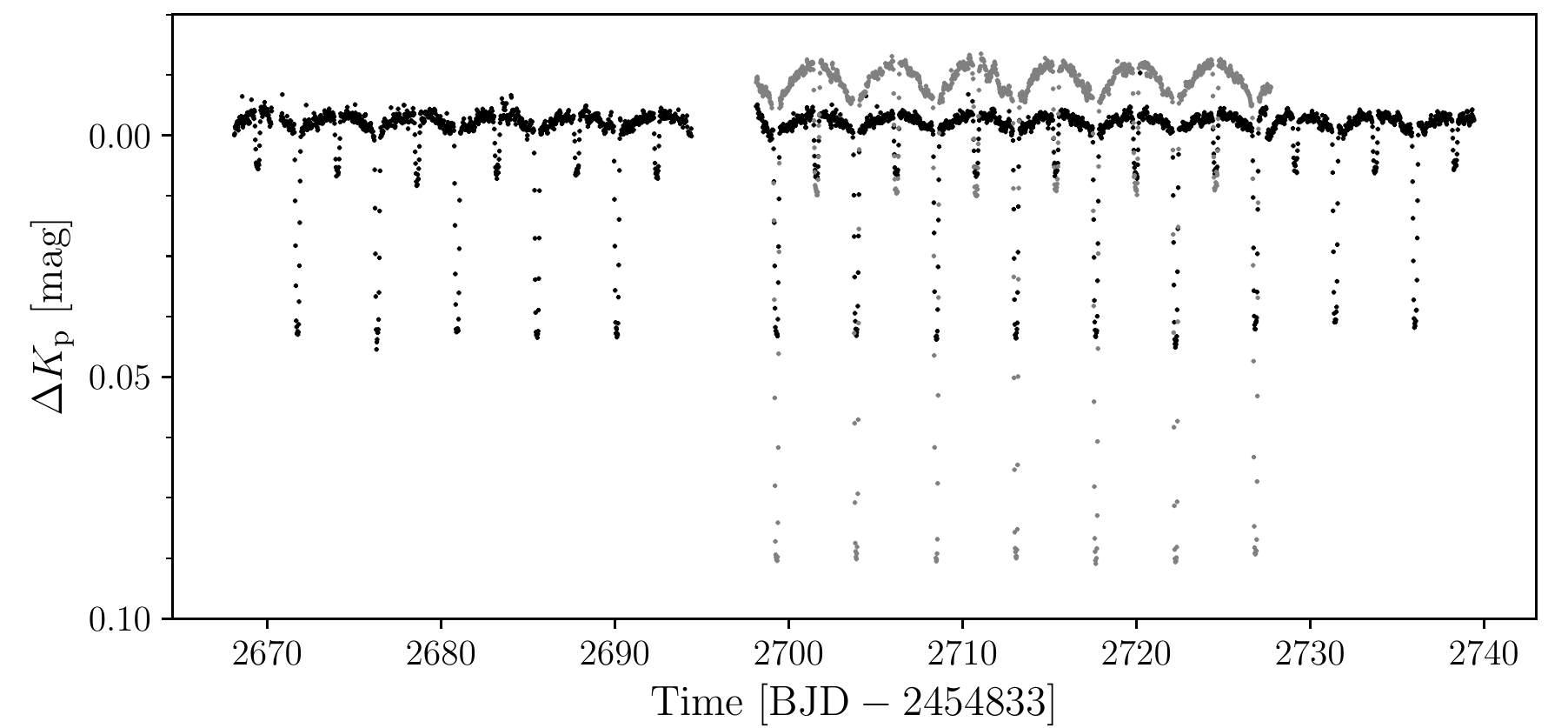}
    \caption{Custom extracted $K2$ light curve of HD~165246 from \citet{Johnston2017}
            ({\sc lc-a}; grey) and halo photometry light curve of HD~165246 ({\sc lc-b}; black).}
    \label{fig:lc_plot}
\end{figure*}

Due to the brightness of HD~165246, the $K2$ pixels saturate during the 30-min-cadence
mode observations. Additionally, HD~165246 lies in a crowded field towards
the galactic centre of the Milky Way and shows potentially contaminating sources
in the surrounding pixels. To address this, \citet{Johnston2017} built a custom pixel
mask that only included the saturated columns and extracted a custom light curve
(hereafter {\sc lc-a}). Due to thruster firings and the saturated and CCD-bleed
columns, however, \citet{Johnston2017} only recovered a $\sim29.9$ d light curve
that was of high enough quality for scientific analysis.

\citet{Johnston2017} concluded that although they identified significant periodicities
consistent with both $p-$ and $g-$mode oscillations, due to the limited time-base of their
light curve and lack of an independent means of identifying the origin of the oscillations {
in combination with the presence of contaminating objects in the pixel image, interpretation
of the signal was limited. To increase the frequency resolution of the light curve}, we extract
a new, longer time-base light curve constructed via the halo photometry method
(hereafter {\sc lc-b}). In this method, the scattered light from bright targets across
the entire pixel image are collected,  with relative weights given to each pixel to scale
their contribution to the resulting light curve \citep{White2017b,Pope2019}. Whereas {\sc lc-a}
{  only has a time-base $T_A\sim29.9$~d, {\sc lc-b} has a time-base of $T_B\sim71.3$~d,
improving the formal frequency resolution from $df_A=1/T_A=0.034$~d$^{-1}$ for {\sc lc-a} to
$df_B=1/T_B=0.014$~d$^{-1}$ for {\sc lc-b}}. However, due to the weighting
scheme used by the halo photometry method, all additional sources present in the mask contribute
contaminating signal to the light curve. This not only increases the overall `third light',
but also potentially adds new variability with an unknown amplitude from a contaminating source.
Since we know neither whether these separate sources are in fact variable, nor
with what amplitudes they may be varying, it is difficult to determine their
individual contributions to {\sc lc-b}. A comparison of the light curves extracted
by \citet{Johnston2017} (grey) and in this work (black) are shown in Fig.~\ref{fig:lc_plot}.
The clear difference in eclipse depth is caused by the large increase in contaminating light
included in {\sc lc-b}.

\subsection{Updated binary model}
\label{section:binary_model}

This section discusses the setup for modelling {\sc lc-a}. The modelling
of {\sc lc-b} is carried out differently, as discussed below. Our initial
binary model is based on the solution of \citet{Johnston2017}, but incorporates
the updated eccentricity and argument of periastron (see
Section~\ref{section:lpv_orbit}), as well as the primary effective temperature
(see Section~\ref{section:atmosphere_soln}). We optimise our binary model using
a Bayesian Markov Chain Monte Carlo (MCMC) numeric sampling code {\tt emcee}
developed by \citet{ForemanMackey2013} and draw uncertainties from the posteriors,
as was done by \citet{Johnston2017}. We calculate the binary model using the
{\tt ellc} code \citep{Maxted2016}. {  In a recent head-to-head comparison, {\tt ellc},
PHOEBE 1 \citep{Prsa2011}, PHOEBE2 \citep{Prsa2016}, JKTEBOP \citep{Southworth2013}
and WD2007 \citep{VanHamme2007} were all shown to agree to $\sim0.4$\% in the reported
fundamental parameters of of the eclipsing binary AI~Phe \citep{Maxted2020}. While the
authors point out that formal uncertainties are often under-estimated when drawn directly
from a co-variance matrix, the consistency of their solutions demonstrate that we should
not expect any systematic offsets in our solution compared to that of \citet{Johnston2017}
based on the use of a different code.} In order to incorporate the spectroscopic
information, we impose Gaussian priors on the eccentricity
$e\sim\mathcal{N}\left(0.029,0.003\right)$, the argument of periastron,
$\omega_0\sim\mathcal{N}\left(1.63,0.09\right)$~rad, the semi-amplitude
of the primary $K_1\sim\mathcal{N}\left(53.0,0.2\right)$~km~s$^{-1}$, the
effective temperature of the primary $T_{\rm eff,1}\sim\mathcal{N}\left(36150,
600\right)$~K, and the projected rotational velocity of the primary
$v_1 \sin i\sim \mathcal{N}\left(268, 25\right)$~km~s$^{-1}$. The
remaining parameters listed in Table~\ref{tab:ellc_soln} are given
uniform priors. We allow the mass ratio, $q=M_2/M_1=K_1/K_2$, to vary
freely so as to not bias the fitting result. Similarly, we allow the third light, $l_3$, to
vary freely as well.

Instead of directly fitting for the effective
temperatures, {\tt ellc} fits for the surface brightness ratio.
To this end, the effective temperature and surface gravities of each
component are only used to interpolate values for the limb-darkening
and gravity-darkening coefficients in the {\it Kepler} passband from
the tables published by \citet{Claret2011}. As such, we sample the
effective temperatures of each component and use the surface gravities
computed for each model as inputs to interpolate for the limb- and gravity
darkening coefficients. Following the suggestion of \citet{Johnston2017},
we include Doppler boosting in our model in order to account for the
asymmetric out-of-eclipse photometric variability observed in the residuals.
We include both the light curve and radial velocity (RV) measurements in
our fitting procedure.

\begin{figure}
    \centering
    \includegraphics[width=0.95\columnwidth]{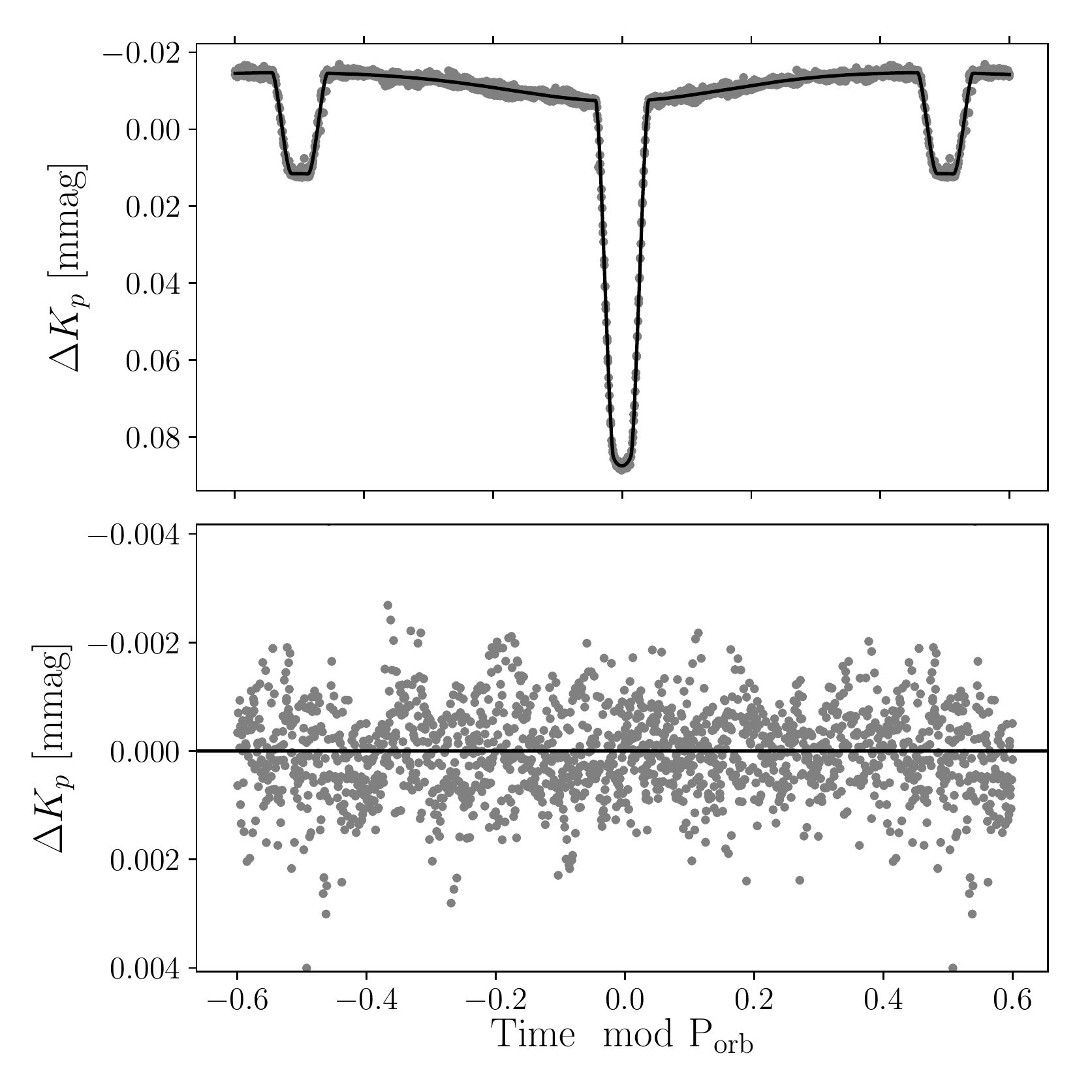}
    \caption{{\bf Top}: $K2$ observations (grey) and best fit model (black) for HD~165246.
             {\bf Bottom}: Residuals (grey) after removal of best fit model.}
    \label{fig:ellc_model}
\end{figure}

We run 10\,000 iterations with 128 individual chains in our MCMC optimisation
routine, discarding those iterations which occurred before five times
the auto-correlation time as burn-in. The parameter estimates and
uncertainties listed in Table~\ref{tab:ellc_soln} are calculated as
the median and 68.3-percentile highest posterior density (HPD) confidence
interval estimates of the marginalised posteriors for each parameter.
The derived quantities, such as the masses and radii of each component,
are calculated at each iteration and saved along with the other sampled
parameters, allowing us to calculate the estimates and uncertainties for
these parameters in the same way. The best fit model, shown in black
in the top panel of Fig.~\ref{fig:ellc_model}, is calculated from the values listed in
Table~\ref{tab:ellc_soln}. The residuals shown in the bottom panel
of Fig.~\ref{fig:ellc_model} have a root-mean-square (RMS) scatter of 0.89 mmag.
We note that the residuals calculated for the same model but without
Doppler boosting have an RMS scatter of 0.91 mmag, and display a
brightening event at $\Phi\sim0.3$, which is consistent with the phase
when the O-star is accelerating towards the line-of-sight. Furthermore,
we note that a significant peak is present at the orbital frequency in the
periodogram of the residuals for the model without beaming included. This
peak is not detected in the residuals of the model with beaming included.

Beyond the inclusion of the eccentricity, boosting factor, and updated
$T_{\rm eff,1}$ in our improved binary model, we find a lower mass ratio compared to
that of \citet{Mayer2013}. We note that this may be caused by the inclusion
of third light in our model. The estimated third light contribution
of $26\substack{+2\\-1}$~per~cent is within the estimates of 7-43~per~cent expected
from the other members of the sextuple system as derived from the K-band
magnitude contrasts published by \citet{Sana2014}. As a combined result,
we calculate that the primary has $M_1=23.7\substack{+1.1\\-1.4}$ M$_{\odot}$
with $R_1=7.3\substack{+0.3\\-0.4}$ R$_{\odot}$, and the secondary has
$M_2$=3$.8\substack{+0.4\\-0.5}$ M$_{\odot}$ with $R_2=2.4\substack{+0.3\\-0.1}$
R$_{\odot}$.

As mentioned previously, the modelling of {\sc lc-b} is conducted differently
to that of {\sc lc-a}. Since we expect the physical binary model to be the same, but
the third light contribution to be different, we fix the model and sample only
the third light contribution for {\sc lc-b}. From this, we find that {\sc lc-b}
has roughly 66~per~cent composite contaminating light, which is significantly
larger than the range expected from the other members of the sextuple system.
This suggests that this light contains rescaled contributions from other stars
in the $K2$ pixel image.

\begin{table}
    \centering
    \caption{Estimated parameters returned from MCMC
             optimised {\tt ellc} binary model. {
             We refer to \citet{Maxted2016} for the
             meaning of the symbols}. Quantities marked
             with $^{\ast}$ were only used to calculate limb-
             and gravity-darkening coefficients.}
    \label{tab:ellc_soln}
    \begin{tabular}{l|c|c}
    \hline
    Parameter & Unit & Estimate \\
    \hline
    \rule{0pt}{2.5ex}$R_1/a$ & -- & $0.208\substack{+0.001 \\-0.001}$ \\
    \rule{0pt}{2.5ex}$R_2/a$ & -- & $0.069\substack{+0.001 \\-0.002}$ \\
    \rule{0pt}{2.5ex}$a$ & $R_{\odot}$ & $35.3\substack{+0.6\\-0.7}$ \\
    \rule{0pt}{2.5ex}$S_b$ & -- & $0.217\substack{+0.004\\-0.006}$ \\
    \rule{0pt}{2.5ex}$q$ & -- & $0.16\substack{+0.02\\-0.02}$ \\
    \rule{0pt}{2.5ex}$i$ & deg & $84.0\substack{+0.1\\-0.1}$ \\
    \rule{0pt}{2.5ex}$P_{\rm orb}$ & d & $4.59270\substack{+0.00001\\-0.00001}$ \\
    \rule{0pt}{2.5ex}$t_0$ & d & $2457215.4108\substack{+0.0004\\-0.0005}$ \\
    \rule{0pt}{2.5ex}$f_c$ & -- & $0.0032\substack{+0.0008\\-0.0008}$ \\
    \rule{0pt}{2.5ex}$f_s$ & -- & $0.1656\substack{+0.0009\\-0.0007}$ \\
    \rule{0pt}{2.5ex}$F_1$ & -- & $3.1\substack{+0.2\\-0.1}$ \\
    \rule{0pt}{2.5ex}$A_1$ & -- & $0.30\substack{+0.1\\-0.2}$ \\
    \rule{0pt}{2.5ex}$B_1$ & -- & $2.0\substack{+0.5\\-0.5}$ \\
    \rule{0pt}{2.5ex}$l_{3,a}$ & \% & $26\substack{+2\\-1}$ \\
    \rule{0pt}{2.5ex}$l_{3,b}$ & \% & $67\substack{+1\\-1}$ \\
    \rule{0pt}{2.5ex}$T_{\rm eff,1}^{\ast}$ & K & $36100\substack{+200\\-200}$ \\
    \rule{0pt}{2.5ex}$T_{\rm eff,2}^{\ast}$ & K & $12100\substack{+600\\-600}$ \\
    \hline
    \multicolumn{3}{c}{Derived Parameters}\\
    \hline
    \rule{0pt}{2.5ex}$e$ & -- & $0.027\substack{+0.003\\-0.002}$ \\
    \rule{0pt}{2.5ex}$\omega_0$ & rad & $1.55\substack{+0.01\\-0.01}$ \\
    \rule{0pt}{2.5ex}$k$ & -- & $0.333\substack{+0.007\\-0.009}$ \\
    \rule{0pt}{2.5ex}$(r_1+r_2)/a$ & -- & $0.2766\substack{+0.002\\-0.001}$ \\
    \rule{0pt}{2.5ex}$L_2/L_1$ & -- & $0.0241\substack{+0.002\\-0.001}$ \\
    \rule{0pt}{2.5ex}$M_1$ & $M_{\odot}$ & $23.7\substack{+1.1\\-1.4}$ \\
    \rule{0pt}{2.5ex}$R_1$ & $R_{\odot}$ & $7.3\substack{+0.3\\-0.4}$ \\
    \rule{0pt}{2.5ex}$\log g_1$ & dex & $4.08\substack{+0.02\\-0.04}$ \\
    \rule{0pt}{2.5ex}$M_2$ & $M_{\odot}$ & { 3}$.8\substack{+0.4\\-0.5}$ \\
    \rule{0pt}{2.5ex}$R_2$ & $R_{\odot}$ & $2.4\substack{+0.3\\-0.1}$ \\
    \rule{0pt}{2.5ex}$\log g_2$ & dex & $4.26\substack{+0.06\\-0.14}$ \\
    \hline
    \end{tabular}
\end{table}

\subsection{Photometric variability}
\label{section:ipw_analysis}

\begin{figure}
    \centering
    \includegraphics[width=0.95\columnwidth]{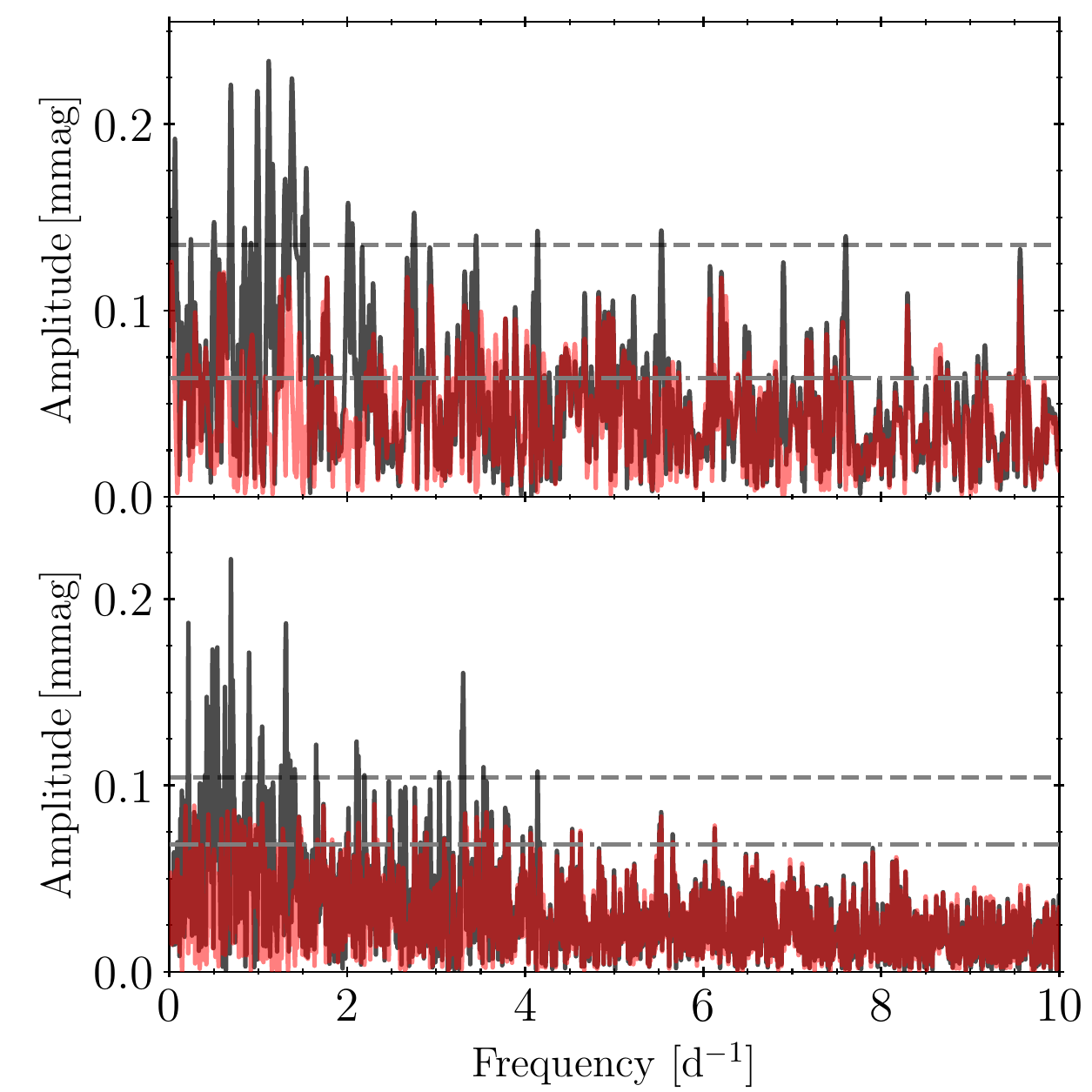}
    \caption{{\bf Top}: Lomb-Scargle Periodogram of {\sc res-a} (black) and its residuals (red).
             {\bf Bottom}: Lomb-Scargle Periodogram of {\sc res-b} (black) and its residuals (red).
             Horizontal dashed dark grey and dashed-dotted light grey lines denote four times
             the white noise level in the residual periodograms after binary model removal and
             prewhitening, respectively.}
    \label{fig:compare_periodograms}
\end{figure}

We use the residuals of {\sc lc-a} and {\sc lc-b} after removal of the
optimised binary model (hereafter {\sc res-a} and {\sc res-b}, respectively)
to calculate a Lomb-Scargle periodogram with the aim of searching for significant
periodicities. The periodograms for {\sc res-a} and {\sc res-b} are shown
in the top and bottom panels of Fig.~\ref{fig:compare_periodograms}. We
subject both {\sc res-a} and {\sc res-b} to an iterative prewhitening process
to extract all variability with an amplitude signal-to-noise ratio (SNR) greater than
four \citep[i.e. SNR>4 ;][]{Breger1993b}, where the SNR is calculated over the full
range $\nu \in $[0, 24.5]~d$^{-1}$ up to the Nyquist frequency. Except for one
frequency, { $f_{14}$}, there is no overlap in the extracted frequency lists
for {\sc res-a} and {\sc res-b}.

Given the greater than factor two increase in contaminating light between the two
light curves, and lack of similarity between the extracted frequency lists from
{\sc res-a} and {\sc res-b}, we only consider those frequencies extracted from
{\sc res-a} in our subsequent analyses. The lack of overlap in extracted
frequencies between the two lists does not imply that the signal is not present
in {\sc res-b}. Rather, given the increase in contaminating light, the signal is
simply no longer significant according to our SNR>4 criterion. We suggest that
the frequencies extracted from {\sc res-b} likely originate in one, or several, of
the contaminating stars included in the halo-photometry mask, and not from the
components of the HD~165246 system. The frequencies extracted from {\sc res-a} are
given in Table~\ref{tab:ipw_results}. The tabulated frequencies have been filtered
for close frequencies occurring within 1.5 times the Rayleigh criterion
\citep{Degroote2009a,Degroote2010,Papics2012,Bowman2017}.

\begin{figure}
    \centering
    \includegraphics[width=0.95\columnwidth]{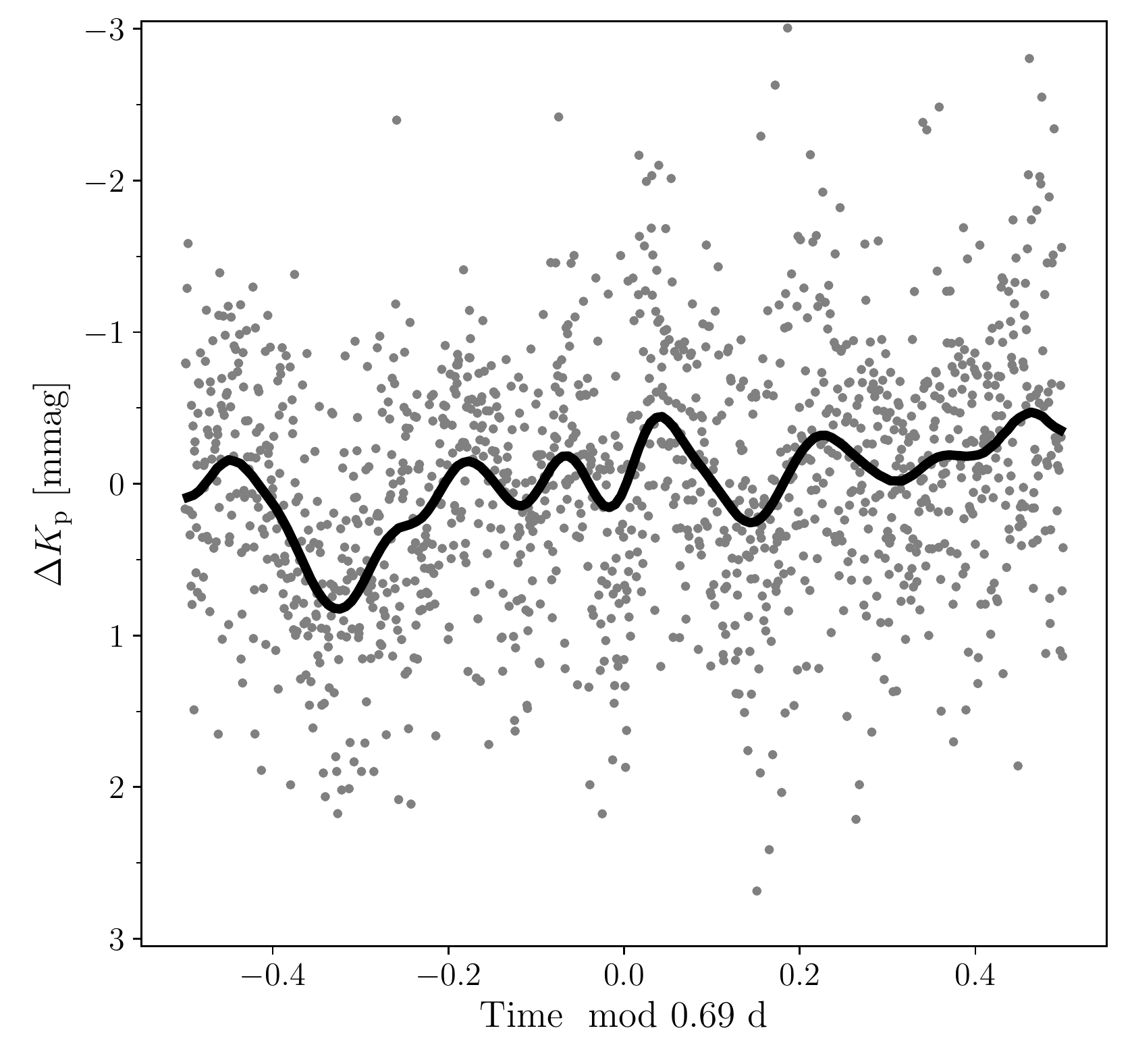}
    \caption{{\sc res-a} phase folded over $f_{\rm p,3}$. Original data in grey, binned data in black.}
    \label{fig:rot_phase}
\end{figure}

In contrast to \citet{Johnston2017}, we identify only one harmonic of
the orbital frequency in {\sc res-a} ($f_{\rm p,5}$). This is a consequence
of the improved binary model. As indicated in Table~\ref{tab:ipw_results},
$f_{\rm p,1}$, $f_{\rm p,2}$, $f_{\rm p,3}$, and $f_{\rm p,7}$ are also
extracted in the LPV analysis in Section~\ref{section:lpv_analysis}. Furthermore,
we find nine components (1,2,3,4,5,6,8,10,11) of a harmonic series with
$f_{\rm p,2}=0.690\pm0.003$~d$^{-1}$ as the base frequency. Extended harmonic series
are the result of non-sinusoidal signals in the light curve, such as binarity,
rotational modulation, or high-amplitude pulsation. The latter option
is excluded as all detected amplitudes are below 1\,mmag.
To investigate the former option, we phase {\sc res-a} over
$f_{\rm p,2}=0.69$ in Fig.~\ref{fig:rot_phase} and find no obvious indication of
a blended binary signal. Assuming $f_{\rm p,2}$ is the rotation frequency, one expects
$F_1=f_{\rm p,2}/f_{\rm orb}=3.17\pm0.01$, which is within { $1-\sigma$} of the
value for $F_1$ obtained from our binary modelling in Section~\ref{section:binary_model},
indicating that the rotational interpretation is feasible. Using $f_{\rm p,2}=f_{\rm rot}$
as well as $R_1$ and $i$ from Table~\ref{tab:ellc_soln} to compute $v_1\sin i$ yields
$v_1\sin i=253^{+11}_{-14}$ km s$^{-1}$. This matches well with the value estimated from
spectroscopy in Section~\ref{section:atmosphere_soln}. This harmonic rotational signal
could be caused by wind variability modulated by the stars rotation (i.e., a clumpy wind),
although we do not detect strong wind signatures in the typical diagnostic lines for
HD~165246. Nevertheless, the { recovery of a significant (SNR>4) signal at}
$f_{\rm s,15}\simeq f_{\rm p,2}$ in the LPVs { with a full 2$\pi$ variation over the
line profile} also suggests an interpretation of this signal in terms of rotational
modulation.

The Lomb-Scargle periodogram of the prewhitened residuals of both {\sc res-a}
and {\sc res-b} are shown in red in the respective panels of Fig.~\ref{fig:compare_periodograms}.
These periodograms showcase stochastic low-frequency variability as found previously for
a large sample of CoRoT, K2, and TESS OB stars \citep{Blomme2011,Aerts2015b,Aerts2018a,SimonDiaz2018,
Bowman2019a,Bowman2019b}. Such a signal is predicted independently by 3D hydrodynamic simulations
carried out by \citet{Edelmann2019} as well as by different 2D hydrodynamic simulations by
\citet{Horst2020} and \citet{Ratnasingam2020}. All of these simulations concerned single,
young stars. However, given the complexity of this multiple system, other physical causes of
this excess may be relevant as well. The stochastic variability occurs in both {\sc res-a}
and {\sc res-b} and is significant according to the  S/N > 4 level when the latter is computed
from the residual periodogram computed from zero frequency up to the Nyquist frequency (see
the dashed-dot horizontal line in Fig.~\ref{fig:compare_periodograms}). Our argument to consider
such a broad range of frequencies to compute the noise level follows \citet{Blomme2011} and
\citet{Bowman2020a}, who have shown that young massive O stars such as the primary component
of HD~165246 have significant low-frequency variability up to frequencies of order 100~d$^{-1}$.

\begin{table}
    \centering
    \caption{Iterative prewhitening results for the residuals of {\sc res-a}.}
    \label{tab:ipw_results}

    \begin{tabular}{lcccc}
    \hline
     & \multicolumn{4}{c}{\sc lc-a}  \\
    ID & {\rm Frequency } & {\rm Amplitude } & {\rm SNR} & Note  \\
       & [d$^{-1}$]       & [mmag]           &           &       \\
    \hline
    $f_{\rm p,1}$ & $0.502\pm0.004$ & 0.15 & 4.61 & $f_{\rm s,9}$ \\
    $f_{\rm p,2}$ & $0.690\pm0.003$ & 0.22 & 6.26 & $f_{\rm s,15}$  \\
    $f_{\rm p,3}$ & $0.992\pm0.003$ & 0.23 & 6.45 & $f_{\rm s,17}$  \\
    $f_{\rm p,4}$ & $1.117\pm0.003$ & 0.23 & 6.55 & -  \\
    $f_{\rm p,5}$ & $1.305\pm0.004$ & 0.15 & 4.59 & 6$f_{\rm orb}$ \\
    $f_{\rm p,6}$ & $1.382\pm0.003$ & 0.21 & 6.20 & 2$f_{\rm p,3}$ \\
    $f_{\rm p,7}$ & $1.502\pm0.004$ & 0.14 & 4.59 & $f_{\rm s,14}$  \\
    $f_{\rm p,8}$ & $1.540\pm0.003$ & 0.18 & 5.22 & -  \\
    $f_{\rm p,9}$ & $2.004\pm0.003$ & 0.16 & 4.92 & -  \\
    $f_{\rm p,10}$ & $2.057\pm0.004$ & 0.15 & 4.68 & 3$f_{\rm p,3}$ \\
    $f_{\rm p,11}$ & $2.165\pm0.004$ & 0.14 & 4.45 & - \\
    $f_{\rm p,12}$ & $2.755\pm0.004$ & 0.16 & 4.70 & 4$f_{\rm p,3}$  \\
    $f_{\rm p,13}$ & $3.448\pm0.004$ & 0.13 & 4.28 & 5$f_{\rm p,3}$ \\
    $f_{\rm p,14}$ & $4.139\pm0.004$ & 0.14 & 4.68 & 6$f_{\rm p,3}$  \\
    $f_{\rm p,15}$ & $5.529\pm0.004$ & 0.15 & 4.64 & 8$f_{\rm 3}$ \\
    $f_{\rm p,16}$ & $6.900\pm0.004$ & 0.15 & 4.16 & 10$f_{\rm p,3}$  \\
    $f_{\rm p,17}$ & $7.601\pm0.004$ & 0.13 & 4.55 & 11$f_{\rm p,3}$ \\
\hline
    \end{tabular}
\end{table}

\section{Spectroscopic Analysis}
\label{section:spectroscopic_analysis}
We also investigate the presence of line-profile variability in HD~165246 on
various time scales. To do this, we obtained 160 observations between 3 May 2017
and 10 October 2019 with the {\sc hermes} spectrograph (R=$85\,000$) \citep{Raskin2011}
attached to the 1.2-m Mercator telescope at El Observatario Roque de los Muchachos in
Santa Cruz de La Palma. As many as 20 consecutive exposures were taken during eight nights
in order to achieve a high temporal resolution. The observations have a mean SNR=85 at
550~nm (ranging from SNR=63 for the lowest quality observation to SNR=112 for the highest
quality observation), with an average integration time of 1200 s, and
are well distributed across the orbit. These observations were subjected to background
 and bias subtraction, flat fielding, wavelength calibration (ThAr lamp spectrum), and
 order merging using the local {\sc hermes} pipeline. The reduced spectra were subsequently
 normalised via spline fitting.

To maximise the SNR for the spectra, we calculate a Least Squares Deconvolved (LSD)
profile \citep{Donati1997,Tkachenko2013}. This method involves convolving a series of
$\delta$ functions of given depths at a given set of wavelengths corresponding to a
pre-determined mask to produce an average line profile from the entire spectrum. Thus,
the expected increase in SNR is proportional to $\sqrt{N}$, where $N$ is the number of
spectral lines used in the mask. Furthermore, by allowing for the simultaneous calculation
of multiple average profiles, the LSD methodology enables the detection of multiple
components in the spectrum. Whereas the spectra of O-stars feature strong {\sc He ii}
lines, the optical spectra of B-type or cooler stars feature strong {\sc He i} or metal
lines, depending on the effective temperature. To this end, the use of different masks
allows for the detection of multiple components should they have a significant light contribution.

\begin{figure}
    \centering
    \includegraphics[width=0.95\columnwidth]{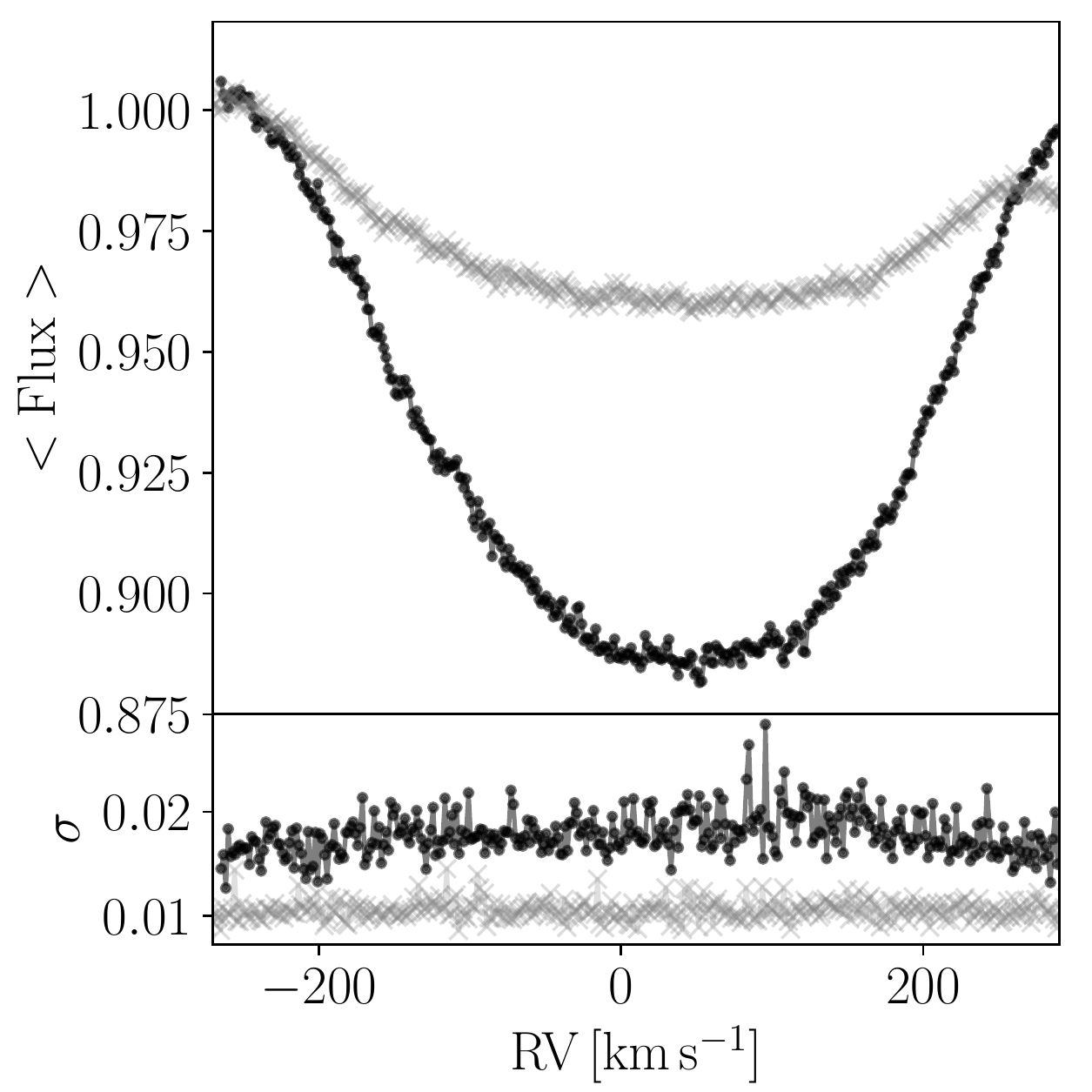}
    \caption{{\bf Top}: Mean LSD profile for mask containing only He lines between 4900 and 5900\AA\ (black) and only
    metal lines across the whole wavelength range (grey). {\bf Bottom}: Standard deviation across LSD profiles.}
    \label{fig:mean_lsds}
\end{figure}

Following these considerations, we subject all of the available {\sc HERMES} spectra to
this method considering two different masks: i) helium lines between 4900 and 5900\AA,
and ii) metal lines, both of which were constructed from the VALD database \citep{Kupka1999}.
Since hydrogen and helium lines are known to suffer from Stark broadening and the signature
of radiation driven winds (should they be present), these lines are generally avoided in line
profile variability studies. However, in the case of hot rapidly rotating stars, helium lines may
be the only lines with sufficient SNR to be considered reliable  \citep{Balona1999,Rivinius2003}.
Figure~\ref{fig:mean_lsds} shows the average LSD profiles constructed using the helium-line
mask (black) and the metal-line mask (grey). The average metal-line profile exhibits a trend
in the red wing. Although it is unclear what introduces this ubiquitous trend in the LSD profiles
produced with the metal-line mask, it is clear that these profiles are unsuitable for line profile
analysis. The average helium-line profile is well behaved. As such, the remainder of the orbital
and line profile analysis is carried out using the LSD profiles produced with the helium-line mask.

\subsection{Line profile variability}

The overall position and shape of a line profile is the combination of extrinsic
and intrinsic broadening effects and perturbations, such as RV shifts due to binarity,
broadening due to rotation, micro- or macroturbulence, and stellar pulsations, some of
which are variable in time. Thus, studying the line profile variations (LPVs) over time
allows for the investigation of these signals.

The velocity field produced by coherent/stochastic stellar pulsations induces
strictly-/quasi-periodic variations in the line-forming regions near the stellar
surface. These variations are detectable via time-resolved spectroscopic observations.
Whereas pressure waves have a predominantly radial contribution to the line profile
\citep{Aerts2003}, gravity waves produce predominantly tangential velocity variations
in the line profile \citep{deCat2002}. It is worth noting that, { although they are
expected to be weak in late O dwarfs}, line-driven winds of massive stars are thought
to be inherently unstable, introducing yet an additional cause of variability into
the line forming region \citep{Puls2008,Sundqvist2011}. However, the observational
consequences of such winds  are only important in cases where they are evident in the
observations. Moreover, wind variability is stochastic and readily distinguishable from
strictly-periodic coherent oscillation modes.

Spectroscopic { frequency} analysis of coherent pulsation modes employs one of two
methods: i) the moment method \citep{Balona1986,Aerts1992,Briquet2003}, or ii) the
pixel-by-pixel method  \citep{Schrijvers1997,Mantegazza2000,Zima2006a,Zima2006b}.
The moment method involves numerically integrating the moments of the extracted spectral
line to describe the variability in terms of the equivalent width (0th moment), the
centroid velocity (corresponding to RV; 1st moment), profile width (2nd moment),
and profile skewness (3rd moment). The moment method is most robust for cases where
the star is not rapidly rotating ($v\sin i<50 \, {\rm km \, s^{-1}}$). There are some
notable exceptions, such as for studying rotational variability in rapidly rotating
chemically peculiar stars \citep[e.g. ][]{Lehmann2006}. However, it
can still be useful for identifying periodicities when combined with the pixel-by-pixel
method for analysis of rapidly-rotating stars. In contrast to the moment method that relies
on the statistical properties of a line profile, the pixel-by-pixel method relies on the
phase and amplitude caused by a stellar pulsation mode across the line profile. While the
pixel-by-pixel method is more useful in cases where $v\sin i\gtrsim50 \, {\rm km \, s^{-1}}$,
it is limited by SNR and is not coupled to the theory of non-radial oscillations as is the case
with the moment method. We use the FAMIAS software package \citep{Zima2008} to carry out the LPV
analysis, using both the moment and pixel-by-pixel methods.

\subsubsection{Orbital variability}
\label{section:lpv_orbit}
\begin{figure}
    \centering
    \includegraphics[width=0.95\columnwidth]{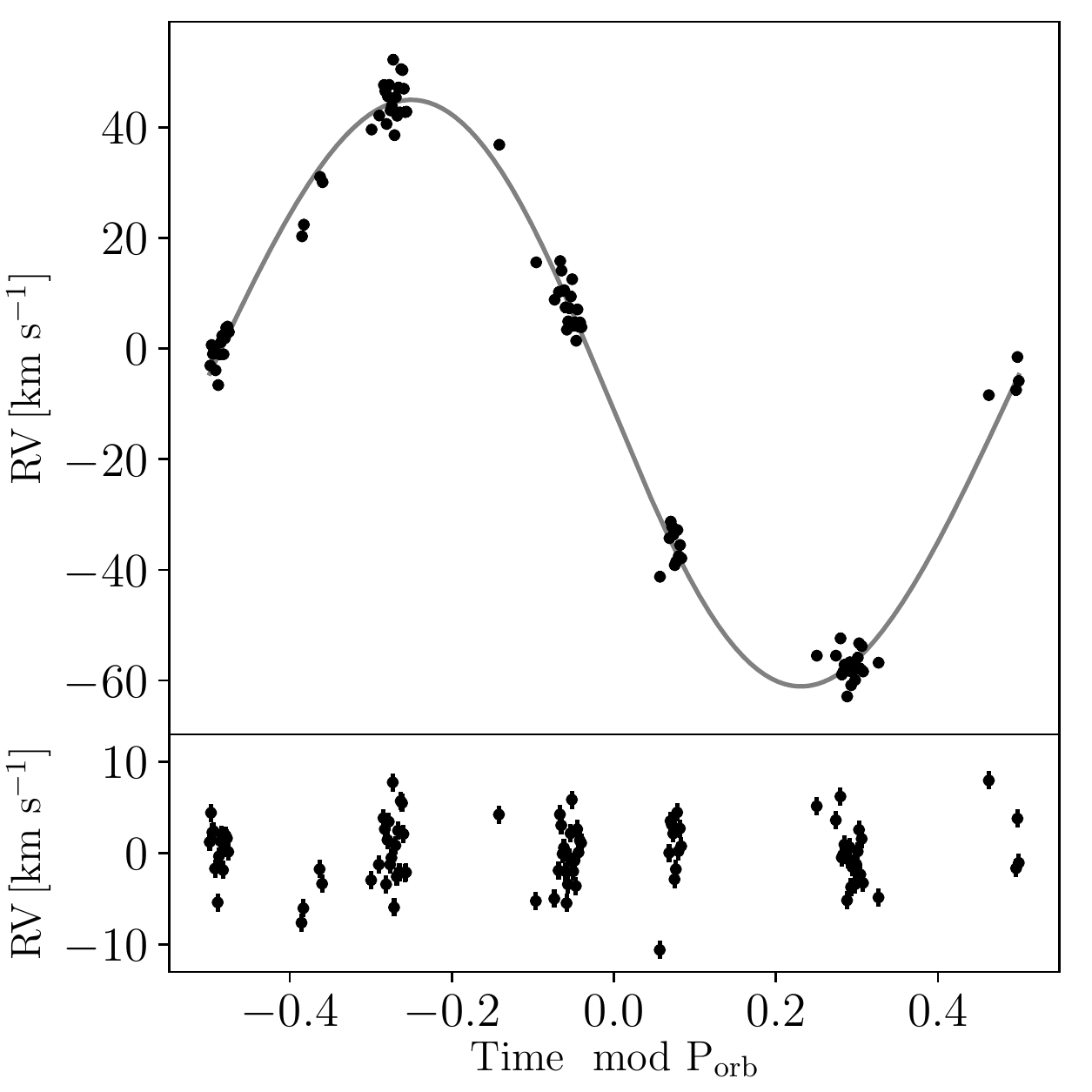}
    \caption{{\bf Top}: Observed RVs for HD~165246A in black, best fit orbital solution
    in grey. {\bf Bottom}: Residuals after subtraction of best fit solution. The observed scatter
    is astrophysical in nature.}
    \label{fig:rv_soln}
\end{figure}

The dominant source of variability among the spectra is the RV shift induced by
the binary motion. In order to investigate any signal caused by stellar pulsations,
we must first effectively model and remove this orbital signal. To do this, we
calculate the first moment for all LSD profiles based on the He mask and fit a
model to these RV shifts using MCMC, as was done with the eclipse modelling in
Section~\ref{section:binary_model}. We fix the orbital period to $P_{\rm orb}=4.59270$~d
and sample the time of periastron passage $t_{\rm pp}$, the eccentricity $e$,
the argument of periastron $\omega_0$, the
semi-amplitude of the primary $K_1$, and the systemic velocity $\gamma$. We derive
estimates and $1\sigma$ uncertainties as the median and 68.3~percentile HPD of the
marginalised posteriors, which are listed in Table~\ref{tab:rv_soln}. The resulting
best fit constructed from these values and the residuals are shown in the top and bottom
panels of Fig.~\ref{fig:rv_soln}. The residuals show a peak-to-peak scatter of $\sim 20$
km s$^{-1}$, indicating the presence of intrinsic variability. Our values for $K_1$,
$\gamma$, $\omega_0$ and $e$ are different to those obtained by \citet{Mayer2013}.
However, given such a small eccentricity and an argument of periastron near 90~deg,
differing solutions for small data sets, such as that used by \citet{Mayer2013}, are
not unexpected.

We do not detect the presence of the secondary in any of the individual spectra,
their LSD profiles, or in the first moment of these profiles. Additionally, we subject
the original data set to spectroscopic disentangling using FDBinary \citep{Ilijic2004,Pavlovski2005},
with a fixed orbital solution according to those values listed in Table~\ref{tab:rv_soln}
and only allow $K_2$ to vary, but are not able to reliably determine a solution.

\begin{table}
    \centering
    \begin{tabular}{lcc}
    \hline
    Parameter & Unit & HPD Estimate \\
    \hline
        \rule{0pt}{2.5ex}$P_{\rm orb}$ & BJD & 4.59270 (fixed)  \\
        \rule{0pt}{2.5ex}$t_{\rm pp}$ & BJD & $2457215.48\substack{+0.06\\-0.06}$ \\
        \rule{0pt}{2.5ex}$e$ & -- & $0.030\substack{+0.003\\-0.003}$ \\
        \rule{0pt}{2.5ex}$\omega_0$ & rad & $1.63\substack{+0.08\\-0.09}$\\
        \rule{0pt}{2.5ex}$K_1$ & ${\rm km\,s^{-1}}$ & $53.0\substack{+0.2 \\ -0.2}$ \\
        \rule{0pt}{2.5ex}$\gamma$ & ${\rm km\,s^{-1}}$ & $-7.9\substack{+0.1\\-0.1}$ \\
    \hline
        \rule{0pt}{2.5ex}$a_1\sin i$ & $R_{\odot}$ & $4.81\substack{+0.01\\-0.01}$\\
        \rule{0pt}{2.5ex}f(M) & $M_{\odot}$ & $0.071\substack{+0.001\\-0.001}$\\
    \hline
    \end{tabular}
    \caption{{\bf Top}: HPD estimates and uncertainties for orbital solution.
    {\bf Bottom}: Derived values { projected distance of the primary to the
    common center of mass and the binary mass function, and their uncertainties.}}
    \label{tab:rv_soln}
\end{table}

\subsubsection{Intrinsic variability}
\label{section:lpv_analysis}

\begin{figure*}
    \centering
    \includegraphics[width=0.98\linewidth]{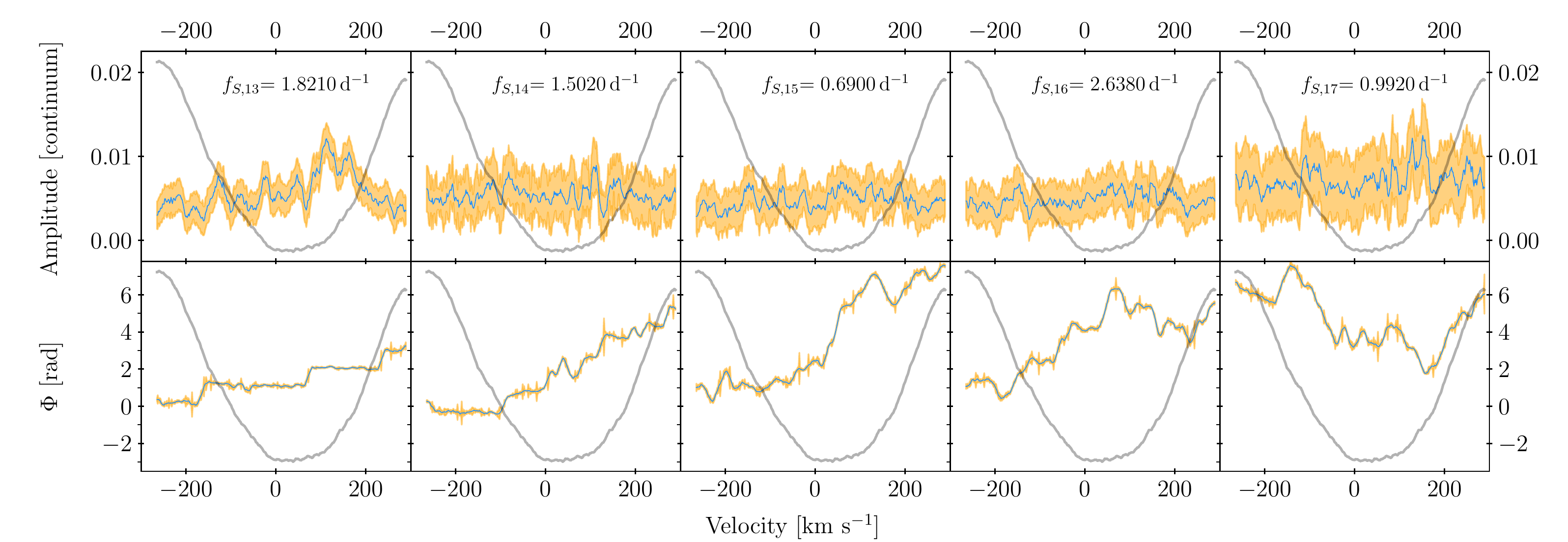}
    \caption{{\bf Top}: Amplitude across line profile. {\bf Bottom}: Phase across line profile. Smoothed
              data displayed in blue, errors displayed in orange. Line profile overplotted in grey in both panels.}
    \label{fig:ppm_results}
\end{figure*}

We remove the orbital motion of the primary from each of the normalised spectra according to
the optimised parameters in Table~\ref{tab:rv_soln}. Following this, we calculate new
LSD profiles using the helium line mask. These LSD profiles are then used in an LPV
analysis, where each LSD profile is assigned a weight according to its SNR. We calculate
the zeroth, first, second, and third moments for the data set and subject them to iterative
prewhitening until all periodicities with SNR>4 are removed from the time-series of these
four moments. We { identify 17} significant frequencies in the { different} moments
and in { the variability across the LSD profiles as found by the so-called velocity
pixel-by-pixel method \citep[see ][]{Zima2008}. In this method one searches for variability
that occur across the entire LSD profile in velocity space, at a given frequency. In
application, we fix the frequency values as found in the photometry or moments and
only retain those frequencies for which variability with a significant (SNR>4) is detected
across the LSD profile.  We fix the frequencies in order to obtain} the highest-precision
fit results for the amplitude and phase behaviour across the LSD profile, as is common practise
in such applications \citep{Zima2006b}. We show the periodograms of the moments in
Appendix~\ref{apdx:pergrams} and the amplitude and phase distributions across the LSDs in
Fig.~\ref{fig:ppm_results}. { We note that 11 of the frequencies recovered from the frequency
analysis of the different moments are newly discovered frequencies, while one is also identified
in the photometry, i.e. $f_{s,9}=f_{p,1}$. The remaining five frequencies recovered via the
pixel-by-pixel method are, as expected, also present in either the photometry or spectroscopy.}

This LPV frequency analysis result is  indicative of low- to high-order p- and g-mode pulsational
variability. Indeed, the frequencies found previously in rapidly rotating $\beta\,$Cep pulsators
are markedly higher than those we find for HD\,165246 \citep[e.g.][]{Schrijvers2004,Uytterhoeven2004,
Uytterhoeven2005}, except for the frequencies $f_{\rm s,4}$  and $f_{\rm s,10.}$ Aside from these two
frequencies, all the other frequencies are lower than those of the p modes found in the CoRoT space photometry
of the slowly rotating O-type dwarf HD\,46202 \citep{Briquet2011}, which is to date the $\beta\,$Cep
star with the highest mass (24\,M$_\odot$) { determined from asteroseismic modelling}.

As expected we find common frequencies among the various { moments} deduced from the
LSD time series. Matches occur between $f_{\rm s,2}$ and $f_{\rm s,11}$, $f_{\rm s,6}$
and $f_{\rm s,13}$, and $f_{\rm s,8}$ and $f_{\rm s,16}$. Additionally, we note that
$f_{\rm s,9}=f_{\rm p,1}$. Of the five frequencies { recovered} by the pixel-by-pixel
method, { three are frequencies from the space photometry and two are from the spectroscopic
moments}. Of these, we note that { $f_{\rm s,15}=f_{\rm p,2}$ is identified as the
rotational frequency.}
Figure~\ref{fig:ppm_results} shows the results of optimising the amplitude (top row) and phase (bottom row)
and their errors, smoothed over a 15~km~s$^{-1}$ window in blue and orange, respectively. Additionally, we
plot the average LSD profile in grey. With the exception of $f_{\rm s,13}$, the amplitudes across the line
profiles are constant within the uncertainties and do not allow us to interpret the results in terms of mode
identification. Phase variability as expected for low-amplitude coherent modes can be seen in some of the
bottom panels of Fig.~\ref{fig:ppm_results}, particularly for $f_{\rm s,13}$. IGWs would result in more
chaotic phase variability across the LSD profiles. The phase variability we detect for $f_{\rm s,14}$,
$f_{\rm s,16}$, and $f_{\rm s,17}$ has a similar level to that found for some of the lowest-amplitude
high-degree p modes found in $\nu\,$Cen \citep{Schrijvers2004}, $\lambda\,$Sco \citep{Uytterhoeven2004},
and $\kappa\,$Sco \citep{Uytterhoeven2005}. In Section~\ref{section:ipw_analysis}, we argued that
$f_{\rm p,2}=f_{\rm s,15}$ can be explained as the rotation frequency of the primary. The full $2\pi$
phase variation across the line profile is consistent with this interpretation.

In conclusion, the complex interplay of frequencies, some of which found in both space photometry and
high-resolution spectroscopy, is not exceptional (e.g., Cotton et al., submitted, treating the high-mass
$\beta\,$Cep pulsator $\beta\,$Cru). This, along with the frequency regime found for the p modes of the
O9V slowly rotating $\beta\,$Cep star HD\,46202 \citep{Briquet2011}, makes us interpret the frequencies
detected in the LSD and in the space photometry of HD\,165246 as due to a mixture of coherent low-order p
and g modes, along with IGWs, shifted into the gravito-inertial regime by the star's fast rotation \citep{Aerts2019}.

\begin{table}
    \centering
    \caption{Significant frequencies, amplitudes, and SNRs extracted from
    0th, 1st, 2nd, and 3rd moments and from the pixel-by-pixel method.}
    \label{tab:lpv_results}

    \begin{tabular}{lcccc}
         \hline
         & d$^{-1}$ & -- & SNR & Note \\
         \hline
        \multicolumn{2}{l}{Zeroeth Moment} & km~s$^{-1}$ &  & \\
        $f_{\rm s,1}$ & $0.18\pm0.02$ & $2.3\pm0.3$ & 21.6 & - \\
        $f_{\rm s,2}$ & $0.487\pm0.004$ & $0.8\pm0.1$ & 8.0 & - \\
        $f_{\rm s,3}$ & $1.729\pm0.003$ & $0.9\pm0.2$ & 8.0 & - \\
        $f_{\rm s,4}$ & $7.342\pm0.002$ & $0.8\pm0.2$ & 4.1 & - \\
         \hline
        \multicolumn{2}{l}{First Moment} & km~s$^{-1}$ &  & \\
        $f_{\rm s,5}$ & $1.476\pm0.002$ & $2.9\pm0.4$ & 7.5 & - \\
        $f_{\rm s,6}$ & $1.821\pm0.002$ & $3.3\pm0.2$ & 8.2 & - \\
         \hline
        \multicolumn{2}{l}{Second Moment} & km$^2$~s$^{-2}$ &  & \\
        $f_{\rm s,7}$ & $2.225\pm0.002$ & $500\pm40$  & 12.6 & - \\
        $f_{\rm s,8}$ & $2.638\pm0.003$ & $300\pm40$  & 7.2 & - \\
        $f_{\rm s,9}$ & $0.501\pm0.003$ & $400\pm100$  & 11.5 & $f_{\rm p,1}$ \\
        $f_{\rm s,10}$ & $4.873\pm0.006$ & $240\pm40$  & 4.5  & - \\
        \hline
        \multicolumn{2}{l}{Third Moment} & km$^3$~s$^{-3}$ &  & \\
        $f_{\rm s,11}$ & $0.489\pm0.002$ & $168000\pm31000$ & 9.6 & - \\
        $f_{\rm s,12}$ & $0.800\pm0.006$ & $150000\pm14000$ & 8.2 & - \\
        \hline
        \multicolumn{2}{l}{Pixel-by-pixel method} & Continuum Units &  &  \\
        $f_{\rm s,13}$ & $1.821$ (fixed) & $3.1\pm0.9$ & 12.4 & $f_{\rm s,6}$ \\
        $f_{\rm s,14}$ & $1.502$ (fixed) & $2.9\pm1.0$ & 6.9 & $f_{\rm p,7}$ \\
        $f_{\rm s,15}$ & $0.690$ (fixed) & $2.7\pm1.0$ & 8.2 & $f_{\rm p,2}$ \\
        $f_{\rm s,16}$ & $2.638$ (fixed) & $2.1\pm1.0$ & 4.8.8 & $f_{\rm s,8}$ \\
        $f_{\rm s,17}$ & $0.992$ (fixed) & $3.9\pm2.0$ & 15.3 & $f_{\rm p,3}$ \\
        \hline
    \end{tabular}
\end{table}

\subsection{Interpretation as variable macroturbulence}
\label{section:macro_var}
It is well known that rotational broadening alone is not sufficient to explain
the shape of line profiles in massive stars \citep{Gray2005}. A microturbulent
velocity component, { which represents turbulent pressure on spatial scales
smaller than the mean-free path of a photon,} is added during the atmosphere
calculations, and can thus alter the line strength { and estimated effective
temperature}. Furthermore, a macroturbulent velocity component, $\xi_{\rm macro}$,
{ which represents turbulent pressure on spatial scales larger than the mean-free
path of a photon,} is required to better reproduce the { shapes of massive star}
line profiles \citep{Howarth1997,Gray2005,Aerts2009,SimonDiaz2014}. { The
macroturbulent velocity profile can be either anisotropic (with different radial and
tangential contributions) or isotropic (with equal radial and tangential contributions).}
Moreover, { while a microturbulent component is considered during the atmospheric
calculations}, both macroturbulence and rotational broadening are included via convolution
to an already computed spectrum, { making their contribution to line shape degenerate.}
Additionally, those stars which undergo  other phenomena such as sub-surface convection,
spots, and/or stellar pulsations which impact the shape of the line profile and make
it asymmetric require further consideration to accurately reproduce the line profile
\citep{Aerts2014b}.

A non-radial pulsation produces asymmetric deviations from the static line profile
that travel through the line profile over the pulsation phase. Thus, the presence of
pulsations can directly influence the measurement of observed quantities, such as
$\xi_{\rm macro}$. \citet{Aerts2009} and \citet{Aerts2015b} demonstrated that the
collective contribution of stellar oscillation modes and IGWs, respectively, can
explain, at least in part, the observed macroturbulence deduced from the spectral-line
properties of massive stars. \citet{Aerts2009,Aerts2014b} also caution that determination
of $v\sin i$ can be complicated by the presence of stellar pulsations which induce
asymmetric time-dependent variations in the line profile, a problem made worse when
spectra obtained at drastically different pulsation phases are stacked. Furthermore,
\citet{Aerts2014b} show that the macroturbulence needed to explain purely pulsational
broadening can be on the order of or larger than the value of the rotational velocity,
and is variable over the pulsation cycle.

As we observe pulsations and a large projected rotational velocity in the O-star primary
of HD~165246, it is important that we obtain an independent estimate of $v\sin i$ to use
as a constraint in the atmospheric modelling. We achieve this by using the {\sc iacob-broad}
tool developed by \citet{SimonDiaz2014} to independently estimate $v\sin i$ for the cases
considering no macroturbulence, using the first moment of the Fourier transform (FT) and
performing goodness-of-fit (GOF) calculations, both allowing for isotropic macroturbulence.
We select 10 spectra from our data set, five of which span the
range of the entire data set and the other five of which were taken on a single night. This
allows us to investigate the stability of the $v\sin i$ and $\xi_{\rm macro}$ estimates over
different timescales and at different points along any variability cycle. Given the SNR of our
spectra, we use the {\sc He ii} 4541 line for our calculations. The results of using
{\sc iacob-broad} on the {\sc He ii} 4541 line are listed in Table~\ref{tab:v_macro}.

As expected, the estimates of $v\sin i$ are systematically higher when $\xi_{\rm macro}$
is fixed at 0 km s$^{-1}$, yielding $v\sin i=268\pm25$~km~s$^{-1}$ compared to
$v\sin i_{\rm FT}= 238\pm40$~km~s$^{-1}$ and $v\sin i_{\rm GOF}= 230\pm46$~km~s$^{-1}$.
The presence of pulsations in HD~165246, however, complicates the
interpretation of this. \citet{Aerts2014b} demonstrated that even low
amplitude pulsations can produce either over- or under-estimations of $v\sin i$, and
hence $\xi_{\rm macro}$, by the FT method if a simple isotropic model of macroturbulence
is used, depending on the pulsation phase of the observed spectrum. This is because a
time-independent isotropic velocity is the wrong prior assumption when fitting profiles
broadened by time-dependent pulsation modes. From this, \citet{Aerts2014b}
conclude, in agreement with \citet{Aerts2009}, that the best means for estimating
$v\sin i$ is via GOF with fixed $\xi_{\rm macro}=0$ km s$^{-1}$. The estimate for
$v\sin i (\xi_{\rm macro}=0)=268\pm25$ km s$^{-1}$ is in agreement with the value for
$v_1\sin i=253^{+11}_{-14}$ km s$^{-1}$ that is calculated assuming that $f_{\rm p,2}$ is the
rotation frequency.

Both the GOF and FT methods of determining  $\xi_{\rm macro}$ reveal that the estimates
of $\xi_{\rm macro}$ are variable on both inter- and intra-nightly timescales, with the
mean estimates exceeding 100~km~s$^{-1}$. This variability in the estimates of $\xi_{\rm macro}$
can be understood in terms of the time-dependent asymmetries produced by pulsations in line
profiles, which require different amounts of isotropic macroturbulence for a satisfactory fit.
Thus, this is not indicative of actual variation in the macroturbulent velocity, but rather the
consequence of measuring $\xi_{\rm macro}$ at different phases of different pulsation cycles.
Furthermore, we recall the 18~km~s$^{-1}$ peak-to-peak scatter observed in the residuals of
the RV fit in Fig.~\ref{fig:rv_soln}. Both the variability in $\xi_{\rm macro}$ and in the RVs
are consistent with the effects of both non-radial coherent p and g modes, as well as IGWs
propagating in the line-forming region of the primary of HD~165246 \citep{Aerts2014b}. Such
modes and waves occur in the frequency range covered by the values listed in
Table~\ref{tab:lpv_results}.

\begin{table*}
    \centering
    \caption{Estimates for $v\sin i$ and $\xi_{\rm macro}$ for 10 spectra.}
    \label{tab:v_macro}
    \begin{tabular}{lccccc}
        \hline
         BJD & $v\sin i \, (\xi_{\rm macro}=0)$ & $v\sin i_{\rm FT}$ & $v\sin i_{\rm GOF}$ & $\xi_{\rm macro,FT}$ & $\xi_{\rm macro,GOF}$ \\
        d & km s$^{-1}$ & km s$^{-1}$ & km s$^{-1}$ & km s$^{-1}$ & km s$^{-1}$  \\
        \hline
        2457896.7177510 & 256 & 235 & 234 & 118 & 117 \\
        2457962.4721318 & 264 & 239 & 237 & 119 & 118 \\
        2457964.5220527 & 282 & 253 & 252 & 136 & 136 \\
        2457965.4924402 & 247 & 225 & 232 & 113 & 113 \\
        2457965.4999635 & 274 & 235 & 213 & 154 & 189 \\
        2457965.5074870 & 265 & 248 & 223 & 120 & 163 \\
        2457965.5150105 & 244 & 217 & 207 & 132 & 153 \\
        2457965.5225342 & 254 & 226 & 224 & 137 & 137 \\
        2457967.4127213 & 333 & 257 & 246 & 222 & 246 \\
        2457971.5365183 & 258 & 241 & 237 & 108 & 108 \\
        \hline
        Mean     & 268    & 238   & 230   & 136   & 148  \\
        Range    & 89     & 40    & 46    & 113   & 137  \\
        $\sigma$ & 25     & 12    & 13    & 31    & 41   \\
        \hline
    \end{tabular}
\end{table*}

\subsection{Updated atmospheric solution}
\label{section:atmosphere_soln}

\begin{figure*}
    \centering
    \includegraphics[width=1.98\columnwidth]{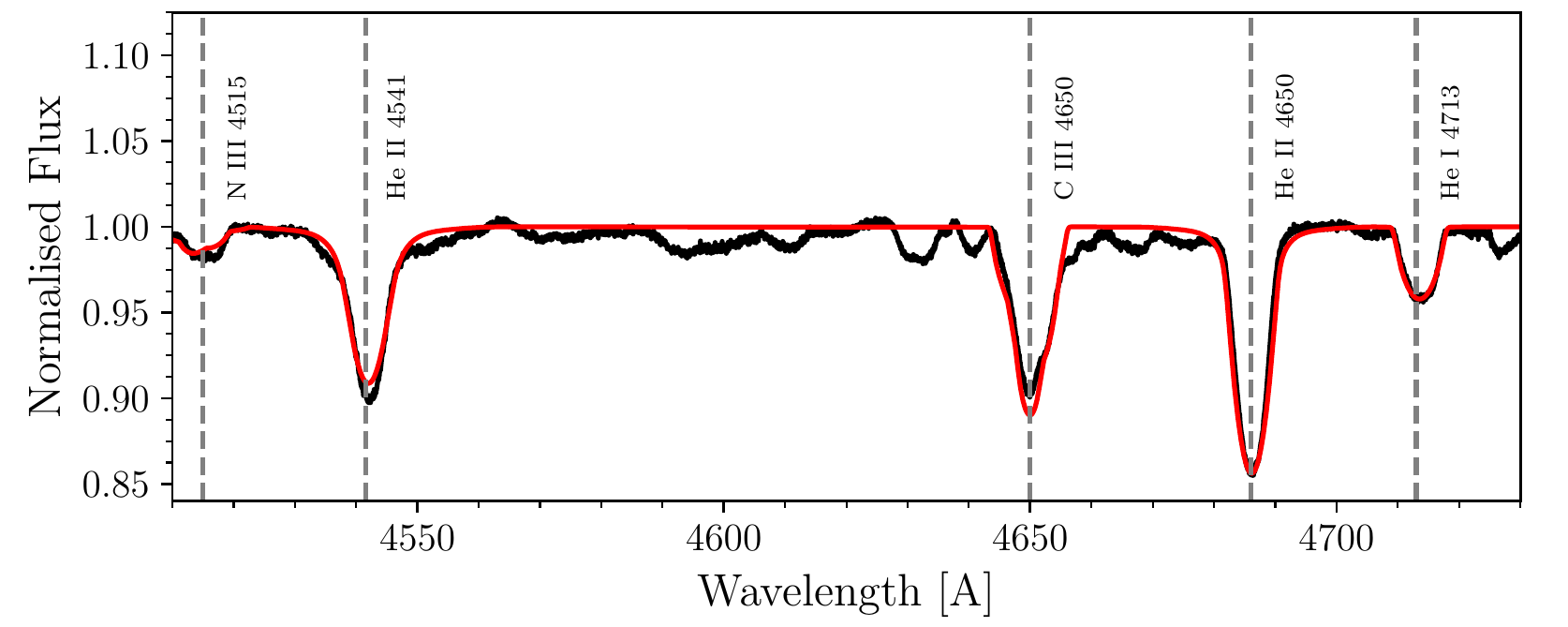}
    \caption{Observed co-added spectrum in black and best fit model according to Table~\ref{tab:spect_soln}
            in red. Location and identification for a subset of spectral lines indicated by vertical dashed lines. }
    \label{fig:fastwind_results}
\end{figure*}

To determine an updated atmospheric solution, we co-add the normalised spectra
(velocity corrected according to the orbital motion), as shown in blue in
Fig.~\ref{fig:fastwind_results}. We perform an atmospheric analysis using the
numerical setup as described in \citet{Abdulmasih2019}. In brief, we use a
genetic algorithm (GA) wrapped around the non-local thermodynamic equilibrium (NLTE)
radiative transfer code {\sc FASTWIND} \citep{Puls2005} to optimise the atmospheric
parameters of the O-star primary \citep{Charbonneau1995, Mokiem2005}. The GA allows
for an efficient sampling of the expansive parameter space and uses a merit function
which is proportional to the inverse of the chi-square of a given atmospheric model
in comparison to a subset of lines from the co-added observed spectrum.

The parameters for each generation of models are determined by combining parameters
from the previous generation, where models with a lower chi-square have a higher
chance of passing their parameters to the next generation. At each generation,
parameter variations, or mutations, are introduced to effectively sample the
parameter space. The GA analysis was carried out iteratively with the binary
modelling (discussed in Section~\ref{section:binary_model}), where the effective
temperature was fixed in the binary modelling first, and then the surface gravity
and light-dilution from the binary model was fixed in the next iteration of spectral
fitting, until convergence was reached.

{ We perform an 11 parameter optimisation including several stellar parameters:
effective temperature ($T_\mathrm{eff}$), surface gravity ($\log g$), microturbulent
velocity ($\xi_{\rm micro}$) and macroturbulent velocity ($\xi_{\rm macro}$), three
wind parameters: the mass loss rate ($\log \Dot{M}$), the exponent of the wind velocity
profile ($\beta$) and the terminal wind speed ($v_\mathrm{inf}$), and four surface
abundance parameters: the helium abundance ($Y_\mathrm{He}$), the carbon abundance
($\eta_\mathrm{C}$), the nitrogen abundance ($\eta_\mathrm{N}$) and the oxygen abundance
($\eta_\mathrm{O}$).  The helium abundance is given as the ratio of the helium number
density to the hydrogen number density while the abundances of carbon, nitrogen and
oxygen are given as the log of the ratio of the elemental number density to the hydrogen
number density plus 12.} The final optimised values for the {\sc FASTWIND} model are listed
in Table~\ref{tab:spect_soln}, and the best fit model is shown in red in
Fig.~\ref{fig:fastwind_results}. Our solution results in a primary effective
temperature which is nearly 3000 K hotter than determined by \citet{Mayer2013}.
Furthermore, we note that we find a high microturbulent velocity in HD~165246,
whereas previous studies typically fix this quantity. Finally, the macroturbulent
velocity reported by the GA optimisation is significantly lower than that obtained
via the GOF and FT methods previously. This is due to differences in the way that
macroturbulence is described, i.e. isotropic in the GOF and FT methods vs.\ anisotropic
in {\sc fastwind}.

\begin{table}
    \centering
    \caption{Estimated parameters returned from optimised FASTWIND models.}
    \label{tab:spect_soln}
    \begin{tabular}{lcc}
    \hline
    Parameter & Unit & Estimate \\
    \hline
    \rule{0pt}{2.5ex}$T_{\rm eff}$ & K    & $36200\substack{+900\\-600}$ \\
    \rule{0pt}{2.5ex}$\log g$  & dex  & $4.05\substack{+0.07\\-0.15}$\\
    \rule{0pt}{2.5ex}$\log \Dot{M}$ & $\log(M_{\odot}/{\rm yr})$ & $-8.0\substack{+0.2\\-0.2}$ \\
    \rule{0pt}{2.5ex}$\beta$ & -- & $0.6787\substack{+0.15\\-0.65}$ \\
    \rule{0pt}{2.5ex}$v_{\infty}$ & ${\rm km\, s^{-1}}$ & $2530\substack{+60\\-260}$ \\
    \rule{0pt}{2.5ex}$\xi_{\rm micro}$ & ${\rm km\, s^{-1}}$ & $13\substack{+1.0\\-1.3}$  \\
    \rule{0pt}{2.5ex}$\xi_{\rm macro}$ & ${\rm km\, s^{-1}}$ & $20\substack{+5\\-6}$ \\
    \rule{0pt}{2.5ex}$v\sin i$ & ${\rm km\, s^{-1}}$ & $268$ (fixed) \\
    \rule{0pt}{2.5ex}$Y_{\rm He}$ & - & $0.092\substack{+0.004\\-0.004}$ \\
    \rule{0pt}{2.5ex}$\eta_{\rm C}$ & - & $8.40\substack{+0.07\\-0.04}$ \\
    \rule{0pt}{2.5ex}$\eta_{\rm N}$ & - & $7.62\substack{+0.1\\-0.09}$ \\
    \rule{0pt}{2.5ex}$\eta_{\rm O}$ & - & $8.67\substack{+0.2\\-0.01}$ \\
    \hline
    \end{tabular}

\end{table}

\section{Discussion}
\label{section:discussion}

We detect multiple sources of variability in both the spectroscopic and photometric
observations of HD~165246. We identify a series of harmonics in the
$K2$ photometry whose base frequency, $f_{\rm p,2}$, is also present in the HERMES
spectroscopy ($f_{\rm p,15}$). The presence of such
harmonics in the photometry indicates the presence of non-sinusoidal signal
in the data, which in this context has two potential explanations: i) rotational
modulation, or ii) background binary signal. Given the corroboration of the binary
modelling, atmospheric modelling, as well as the identification of this signal in
both the photometric and spectroscopic time series, we argue that this is the result
of rotational modulation on the primary O8~V star.

Assigning a single underlying mechanism to the remaining variance in the low-frequency
regime is challenging since it contains both coherent p and g modes self-excited via the
$\kappa$-mechanism as well as stochastically excited IGWs, which may also drive modes at resonant
eigenfrequencies \citep{Bowman2019a,Edelmann2019,Horst2020}. Given the poor frequency
resolution of both the spectroscopic and photometric data sets, we are currently unable
to identify the modes/waves. Since we do, however, identify signal in both the photometry and
spectroscopy independently, we are able to conclusively state that there is significant
pulsational variability present at low frequencies which  originates from the O-type primary.
Moreover, the dominant frequencies detected in the moment variations listed in Table\,\ref{tab:lpv_results}
all lead to a ratio of the tangential to radial velocity amplitude above unity. This ratio,
{ otherwise known as a K value,} can be computed from the mass, radius and frequency of
the mode following Eq.\,(3.162) in \citet{Aerts2010}. This leads to ratios with a range covering
roughly $[3.3,62.2]$, meaning that the pulsational variability is dominated by g modes or IGWs as
these have dominant tangential motions, while p modes are dominated by radial motions. The signal
corresponding to $f_{\rm s,4}$ and $f_{\rm s,10}$ leads to $low$ K values and may correspond to
either low-order p modes \citep{Briquet2011} or high-order g modes shifted to high frequencies due
to the Coriolis force \citep{Buysschaert2018b}.

The observed $\xi_{\rm macro}$ variability may have various candidate sources:
i) coherent pulsations, ii) IGWs,  iii) sub-surface convection, or iv) stochastic
wind variability. As we have identified the presence of pulsations and/or IGWs, these
invariably have at least some contribution to the variability in $\xi_{\rm macro}$
on the basis of their contribution to the line profiles from which $\xi_{\rm macro}$
is estimated. The parameters of the O8~V primary place it in a region on the HR diagram
where the sub-surface convective velocity is theoretically estimated to be below
2.5~km~s$^{-1}$ \citep[][see their Fig.~9, top panel]{Cantiello2009}. Therefore,
sub-surface convection cannot fully explain the large and variable tangential velocities
that we observe. Furthermore, our estimates of $\xi_{\rm micro}$, $\log g$, and
$v\sin i$ place the O-star primary in an underpopulated region of the parameter
space to compare with the predictions of \citet{Cantiello2009}. This comparison,
however, is complicated by the fact that HD~165246 is within the galaxy, whereas the
majority of the sample analysed by \citet{Cantiello2009} consists of stars from the Large
and Small Magellanic Clouds. In addition, stochastically variable wind signatures as
computed by \citet{Krticka2018} stem from outflow and hence correspond dominantly
to radial motions, while the observations point to dominant tangential velocity
variations. Our observations of high $v\sin i$, high $\xi_{\rm macro}$, and the
presence of IGWs in the young O-star primary of HD~165246 are consistent with the
results of both \citet{SimonDiaz2017} and \citet{Bowman2019b,Bowman2020a}.
\citet{SimonDiaz2017} observe a wide range of $\xi_{\rm macro}$ from spectroscopy
and \citet{Bowman2019b} observe a low-frequency excess in photometry (identified as
IGWs) in both galactic and LMC O and B stars across the upper HRD. This suggests a
common intrinsic mechanism and a relationship between macroturbulence as found in
spectroscopy and stochastic low-frequency variability detected in space photometry
\citep{Burssens2020,Bowman2020a}.

\begin{figure}
    \centering
    \includegraphics[width=0.95\columnwidth]{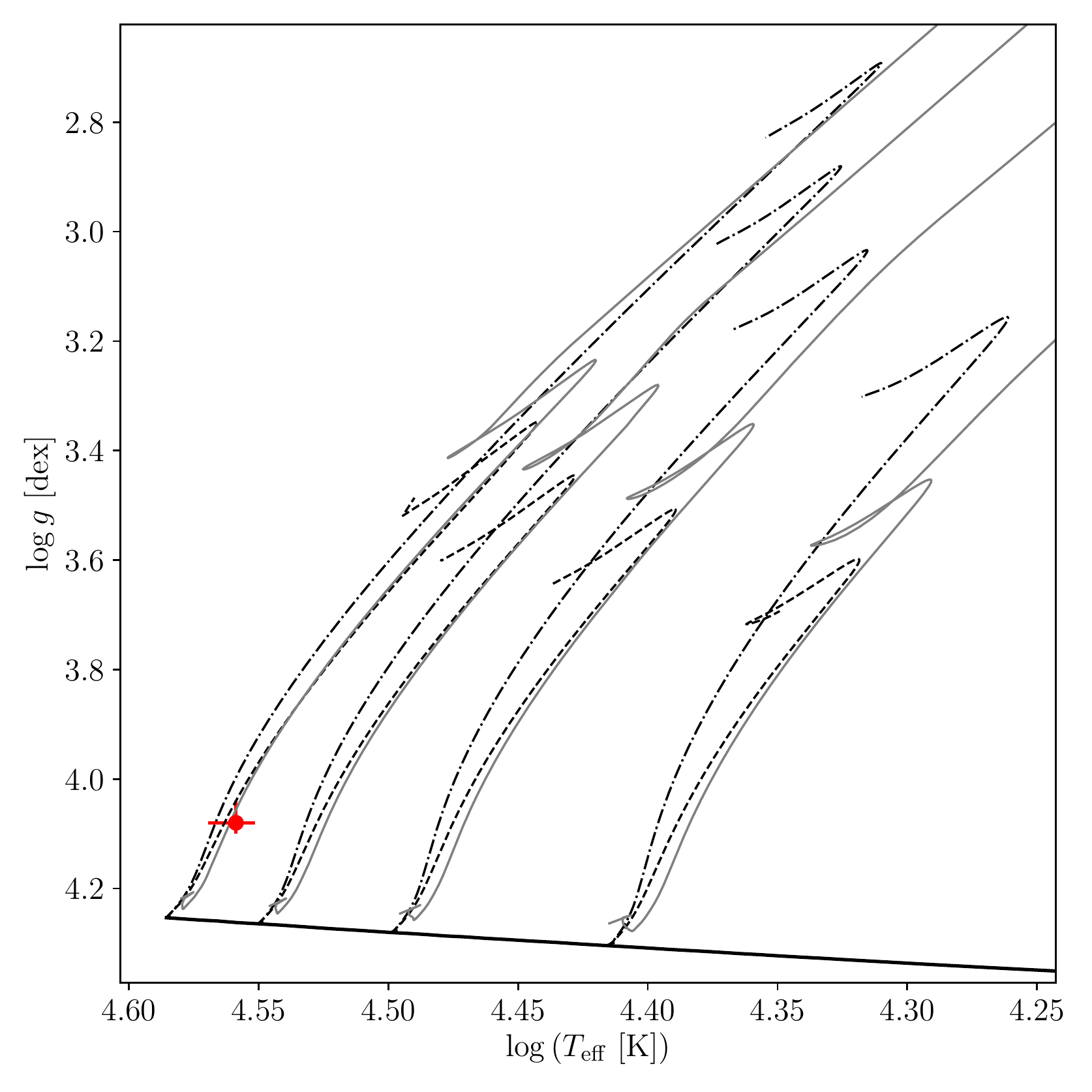}
    \caption{ Evolutionary tracks for 25, 20, 15, and 10 M$_{\odot}$ models. Solid black
    line denotes ZAMS line. Black tracks taken from \citet{Johnston2019a} with $f_{\rm ov}=0.005$,
    $D_{\rm REM}=10$\,cm$^2$\,s$^{-1}$ (dashed lines) and $f_{\rm ov}=0.040$,
    $D_{\rm REM}=10\,000$\,cm$^2$\,s$^{-1}$ (dashed-dotted lines). { Black tracks have Y=0.276, Z=0.0140,
    and use OP opacities. Solid grey tracks are MIST Stellar evolution tracks taken from \citet{Choi2016}
    with $v_{\rm ZAMS}/v_{\rm crit}=$ 0.4, Y=0.2703, Z=0.0142, and OPAL opacities, for the same masses.} }
    \label{fig:kiel}
\end{figure}

As demonstrated in Fig.~\ref{fig:kiel}, the O8~V primary is located close to the
zero-age main-sequence in reference to both the non-rotating tracks with different amounts
on internal chemical mixing (black tracks), calculated according to \citet{Johnston2019a}
using {\sc mesa} \citep[r-10398][]{Paxton2018} as well as tracks with { an initial
rotational velocity of 40\% critical (grey tracks), calculated by \citet{Choi2016} using
{\sc mesa} \citep[r-7503][]{Paxton2015}.}
The non-rotating models are calculated such that the internal chemical element mixing
is represented in two distinct regimes. The first regime corresponds to the convective
boundary mixing (CBM) region, { where we represent CBM with diffusive convective
overshooting. For overshooting}, the free parameter $f_{\rm ov}$ scales the slope with
which the mixing profile decays beyond the core, as defined by the Schwarzschild criterion
in terms of the local pressure scale height, $H_p$. { These models only account for
overshooting extending beyond the convective core, and not around any intermediate convective
regions in the envelope.} The second regime corresponds to radiative envelope mixing (REM),
according to the { chemical mixing induced by internal gravity waves as derived by
\citet{Rogers2017} and} implemented { in MESA models} by \citet{Pedersen2018}. In this profile, the free
parameter $D_{\rm REM}$ sets the base efficiency of chemical mixing induced by this
mechanism, and has units of cm$^2$~s$^{-1}$. The dashed black (non-rotating) models plotted
in Fig.~\ref{fig:kiel} represent the case of minimal internal chemical mixing, with
$f_{\rm ov}=0.005$ and $D_{\rm REM}=10$\,cm$^2$\,s$^{-1}$. The dashed-dotted black
(non-rotating) models represent the case of maximum internal mixing with $f_{\rm ov}=0.04$
and $D_{\rm REM}=10\,000$\,cm$^2$\,s$^{-1}$. These limiting values are
those deduced by asteroseismology of intermediate-mass g-mode pulsating field stars
\citep{Briquet2007,Daszynska2010,Daszynska2013,Moravveji2016,Schmid2016,Buysschaert2018b,
Walczak2019,Wu2019}. {  Despite the differences in the form of internal mixing between
the two sets of evolutionary tracks, comparison of the observations with both tracks
reveals a massive primary which has consumed less than 30\% of its initial core hydrogen
content, as seen in Fig.~\ref{fig:kiel}.}

\begin{figure}
    \centering
    \includegraphics[width=0.95\columnwidth]{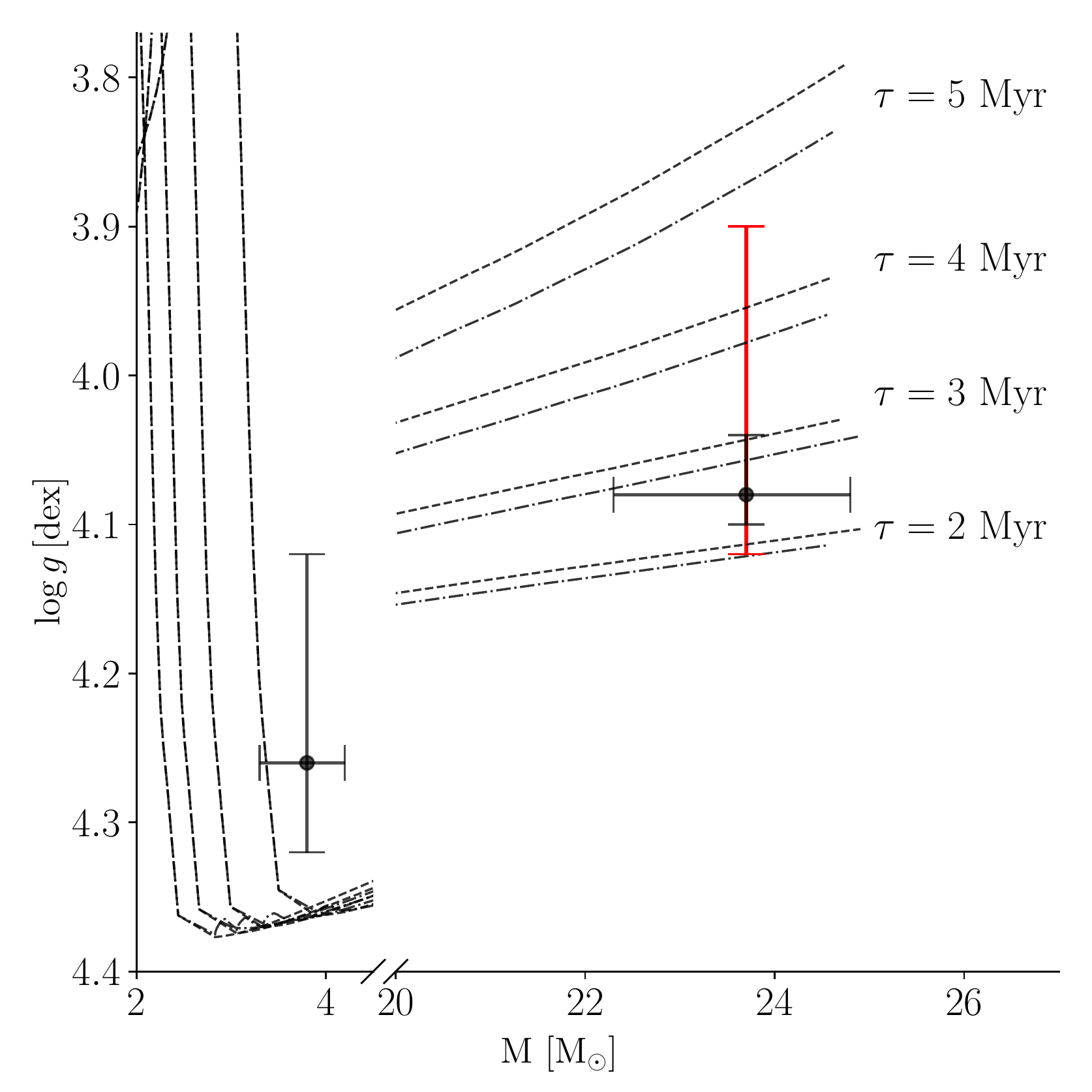}
    \caption{ Isochrones for 2,3,4, and 5 Myr shown in grey for tracks with
             $f_{\rm ov}=0.005$, $D_{\rm REM}=10$\,cm$^2$\,s$^{-1}$ (dashed lines) and
             $f_{\rm ov}=0.040$, $D_{\rm REM}=10\,000$\,cm$^2$\,s$^{-1}$ (dashed-dotted lines).
             All models have Y=0.276 and Z=0.014 and OPAL opacities.
             Dynamical (spectroscopic) estimates for mass and $\log g$ denoted by
             black (red) error-bars.
             }
    \label{fig:hrd}
\end{figure}

Figure~\ref{fig:hrd} compares the dynamical mass and surface gravity estimates
for the O8V primary of HD~165246 with theoretical isochrones with different
amounts of internal mixing using models calculated using {\sc mesa} \citep{Paxton2018}
by \citet{Johnston2019a}. The location of HD~165246A in Fig.~\ref{fig:hrd} indicates
that it has an age between 2-3~Myr and a core hydrogen content of $X_c=0.54$, given the
uncertainties on the dynamical surface gravity (black), or an age between 2-4.5~Myr
given the uncertainties on the spectroscopic surface gravity (red). Our age estimates
agree with those of \citet{Mayer2013} who used evolutionary tracks from \citet{Brott2011a}.
The overlap between the dynamical and spectroscopic surface gravity estimates provides
support for the mass estimate derived for the primary component. However, given the inability
to detect the secondary star in the spectra, we have no means of independently verifying the
solution for the secondary, making us wary of its absolute dimensions. With this in mind, we
find that the secondary is not yet on the main-sequence when compared to the isochrones
in Fig.~\ref{fig:hrd}. We note that without smaller uncertainties on the parameters of the
primary, we are not able to constrain the impact of internal chemical mixing on the evolution of
this star. Additionally, without proper characterization of the secondary via detection of
its RV variations, we are not able to perform isochrone-cloud fitting as introduced by
\citet{Johnston2019b} and applied by \citet{Tkachenko2020}.

\section{Conclusions}
\label{section:conclusions}

This work made use of an extensive set of new time-series spectroscopy to
build upon the work by \citet{Mayer2013} and \citet{Johnston2017} for a more
in-depth study of the massive eclipsing O+B binary HD~165246. Using this spectroscopy,
we obtained updated atmospheric and binary solutions revealing an effective temperature
of $T_{\rm eff} = 36\,000$~K and a mass of $M = 23.7$~M$_{\odot}$ for the O-star
primary.  Furthermore, we determine a rotation rate of
$v\sin i=268$~km~s$^{-1}$, microturbulence of $\xi_{\rm micro}=13$~km~s$^{-1}$, and
variable macroturbulence linked to time-dependent pulsational line-broadening.
We explored the variability detected in both the $K2$ photometry and {\sc hermes}
spectroscopy, and find it consistent with rotational modulation, as well as
intrinsically excited stellar pulsations.

The presence of pulsational variability in the light curve and {\sc hermes}
spectroscopy provide at least a partial explanation for the high macroturbulence
derived from the atmospheric modelling. This is consistent with the
predictions of \citet{Aerts2009, Aerts2014b} and the observations of
\citet{SimonDiaz2017}. The evolutionary status of the O8V primary
is consistent with expectations from models. Higher precision
on the fundamental stellar parameters is required for scrutinising
evolutionary models at masses and ages typical of O-stars. However,
the observed atmospheric properties of the primary component situate this star
in an underpopulated region of the parameter space of massive star
variability studies.

High-mass eclipsing binaries such as HD~165246 are integral to the
study of variability in the upper regions of the HRD. The characterization
of such systems is required to provide benchmarks for hydrodynamical simulations
and evolutionary codes. Currently, the community lacks the sample
size of well characterized high-mass stars (including fundamental parameters)
required to discriminate between the physical mechanisms proposed as
the causes of observed variability. The Transiting Exoplanet Survey Satellite
\citep[TESS;][]{Ricker2015} is currently assembling sub-mmag precision
space-based photometry of high-mass stars across the entire sky. This sample
is yielding high quality observations of more massive stars than any homogeneous
database to date \citep[see e.g.,][]{Burssens2020,LabadieBartz2020}. Characterization of the
variability (or lack thereof) in these stars is providing the basis upon which
predictions of physical mechanisms can be tested. In particular, stars that
reveal coherent oscillation modes whose degree and azimuthal order can be identified,
are suitable to be probeb via asteroseismology \citep{Aerts2020,Bowman2020b}. So far,
such detections and identifications are typically achieved in only about 10\% of
the observed OB-type stars, but this fraction is expected to increase thanks to TESS.

\section*{Data Availability}

The K2 data presented in this paper were obtained from the Mikulski Archive for Space
Telescopes (MAST) at the Space Telescope Science Institute (STScI), which is operated
by the Association of Universities for Research in Astronomy, Inc., under NASA contract
NAS5-26555. Support to MAST for these data is provided by the NASA Office of Space Science
via grant NAG5-7584 and by other grants and contracts. Funding for the Kepler/K2 mission
was provided by NASA’s Science Mission Directorate. The {\sc hermes} data used within this
article are available upon request. The binary modelling
was carried out using {\sc ellc} \citep{Maxted2016} and the atmospheric modelling was carried
out using {\sc Fastwind} \citep{Puls2005}. The propagation of asymmetric uncertainties was
performed using the {\sc SOAD} code \citep{Erdim2019}. The optimisation routines in this
work made use of the {\sc Numpy} \citep{Harris2020} and {\sc emcee} \citep{ForemanMackey2013},
and the plotting was carried out using {\sc Matplotlib} \citep{Hunter2007}. Mode identification
results obtained with the software package FAMIAS developed in the framework of the
FP6 European Coordination Action HELAS (http://www.helas-eu.org/).

\section*{Acknowledgements}
We thank the referee for their comments which improved the manuscript.
The research leading to these results has received funding from the European
Research Council (ERC) under the European Union's Horizon 2020 research and
innovation programme (grant agreement N$^\circ$670519: MAMSIE and grant
agreement N$^\circ$772225: Multiples), from the KU\,Leuven Research Council
(grant C16/18/005: PARADISE and C16/17/007: MAESTRO), from the Research
Foundation Flanders (FWO) under grant agreements G0H5416N (ERC Runner Up
Project) and G0A2917N (BlackGEM), from the FWO-Odysseus program under
G0F8H6N, as well as from the BELgian federal Science Policy Office (BELSPO)
through PRODEX grant PLATO. DMB gratefully acknowledges a senior postdoctoral
fellowship from the Research Foundation Flanders (FWO) with grant agreement
No. 1286521N. The computational resources and services used in this work were
provided by the VSC (Flemish Supercomputer Center), funded by the
Research Foundation - Flanders (FWO) and the Flemish Government -
department EWI to PI Johnston.




\bibliographystyle{mnras}
\bibliography{astroBib.bib} 

\begin{thebibliography}{}
\makeatletter
\relax
\def\mn@urlcharsother{\let\do\@makeother \do\$\do\&\do\#\do\^\do\_\do\%\do\~}
\def\mn@doi{\begingroup\mn@urlcharsother \@ifnextchar [ {\mn@doi@}
  {\mn@doi@[]}}
\def\mn@doi@[#1]#2{\def\@tempa{#1}\ifx\@tempa\@empty \href
  {http://dx.doi.org/#2} {doi:#2}\else \href {http://dx.doi.org/#2} {#1}\fi
  \endgroup}
\def\mn@eprint#1#2{\mn@eprint@#1:#2::\@nil}
\def\mn@eprint@arXiv#1{\href {http://arxiv.org/abs/#1} {{\tt arXiv:#1}}}
\def\mn@eprint@dblp#1{\href {http://dblp.uni-trier.de/rec/bibtex/#1.xml}
  {dblp:#1}}
\def\mn@eprint@#1:#2:#3:#4\@nil{\def\@tempa {#1}\def\@tempb {#2}\def\@tempc
  {#3}\ifx \@tempc \@empty \let \@tempc \@tempb \let \@tempb \@tempa \fi \ifx
  \@tempb \@empty \def\@tempb {arXiv}\fi \@ifundefined
  {mn@eprint@\@tempb}{\@tempb:\@tempc}{\expandafter \expandafter \csname
  mn@eprint@\@tempb\endcsname \expandafter{\@tempc}}}

\bibitem[\protect\citeauthoryear{{Abdul-Masih} et~al.,}{{Abdul-Masih}
  et~al.}{2019}]{Abdulmasih2019}
{Abdul-Masih} M.,  et~al., 2019, \mn@doi [\apj] {10.3847/1538-4357/ab24d4},
  \href {https://ui.adsabs.harvard.edu/abs/2019ApJ...880..115A} {880, 115}

\bibitem[\protect\citeauthoryear{{Aerts}}{{Aerts}}{2021}]{Aerts2020}
{Aerts} C.,  2021, RMP, in press, \href
  {https://ui.adsabs.harvard.edu/abs/2019arXiv191212300A} {39,
  arXiv:1912.12300}

\bibitem[\protect\citeauthoryear{{Aerts} \& {De Cat}}{{Aerts} \& {De
  Cat}}{2003}]{Aerts2003}
{Aerts} C.,  {De Cat} P.,  2003, \mn@doi [\ssr] {10.1023/A:1023983704925},
  \href {https://ui.adsabs.harvard.edu/abs/2003SSRv..105..453A} {105, 453}

\bibitem[\protect\citeauthoryear{{Aerts} \& {Rogers}}{{Aerts} \&
  {Rogers}}{2015}]{Aerts2015b}
{Aerts} C.,  {Rogers} T.~M.,  2015, \mn@doi [\apjl]
  {10.1088/2041-8205/806/2/L33}, \href
  {https://ui.adsabs.harvard.edu/abs/2015ApJ...806L..33A} {806, L33}

\bibitem[\protect\citeauthoryear{{Aerts}, {de Pauw}  \& {Waelkens}}{{Aerts}
  et~al.}{1992}]{Aerts1992}
{Aerts} C.,  {de Pauw} M.,   {Waelkens} C.,  1992, \aap, \href
  {https://ui.adsabs.harvard.edu/abs/1992A&A...266..294A} {266, 294}

\bibitem[\protect\citeauthoryear{{Aerts}, {Puls}, {Godart}  \&
  {Dupret}}{{Aerts} et~al.}{2009}]{Aerts2009}
{Aerts} C.,  {Puls} J.,  {Godart} M.,   {Dupret} M.~A.,  2009, \mn@doi [\aap]
  {10.1051/0004-6361/200810471}, \href
  {https://ui.adsabs.harvard.edu/abs/2009A&A...508..409A} {508, 409}

\bibitem[\protect\citeauthoryear{{Aerts}, {Christensen-Dalsgaard}  \&
  {Kurtz}}{{Aerts} et~al.}{2010}]{Aerts2010}
{Aerts} C.,  {Christensen-Dalsgaard} J.,   {Kurtz} D.~W.,  2010,
  {Asteroseismology, Astronomy and Astrophysics Library}.
Springer-Verlag, Heidelberg

\bibitem[\protect\citeauthoryear{{Aerts}, {Sim{\'o}n-D{\'\i}az}, {Groot}  \&
  {Degroote}}{{Aerts} et~al.}{2014}]{Aerts2014b}
{Aerts} C.,  {Sim{\'o}n-D{\'\i}az} S.,  {Groot} P.~J.,   {Degroote} P.,  2014,
  \mn@doi [\aap] {10.1051/0004-6361/201424012}, \href
  {https://ui.adsabs.harvard.edu/abs/2014A&A...569A.118A} {569, A118}

\bibitem[\protect\citeauthoryear{{Aerts}, {Van Reeth}  \& {Tkachenko}}{{Aerts}
  et~al.}{2017}]{Aerts2017}
{Aerts} C.,  {Van Reeth} T.,   {Tkachenko} A.,  2017, \mn@doi [\apjl]
  {10.3847/2041-8213/aa8a62}, \href
  {https://ui.adsabs.harvard.edu/abs/2017ApJ...847L...7A} {847, L7}

\bibitem[\protect\citeauthoryear{{Aerts} et~al.,}{{Aerts}
  et~al.}{2018}]{Aerts2018a}
{Aerts} C.,  et~al., 2018, \mn@doi [\mnras] {10.1093/mnras/sty308}, \href
  {https://ui.adsabs.harvard.edu/abs/2018MNRAS.476.1234A} {476, 1234}

\bibitem[\protect\citeauthoryear{{Aerts}, {Mathis}  \& {Rogers}}{{Aerts}
  et~al.}{2019}]{Aerts2019}
{Aerts} C.,  {Mathis} S.,   {Rogers} T.~M.,  2019, \mn@doi [\araa]
  {10.1146/annurev-astro-091918-104359}, \href
  {https://ui.adsabs.harvard.edu/abs/2019ARA&A..57...35A} {57, 35}

\bibitem[\protect\citeauthoryear{{Aldoretta} et~al.,}{{Aldoretta}
  et~al.}{2015}]{Aldoretta2015}
{Aldoretta} E.~J.,  et~al., 2015, \mn@doi [\aj] {10.1088/0004-6256/149/1/26},
  \href {https://ui.adsabs.harvard.edu/abs/2015AJ....149...26A} {149, 26}

\bibitem[\protect\citeauthoryear{{Balona}}{{Balona}}{1986}]{Balona1986}
{Balona} L.~A.,  1986, \mn@doi [\mnras] {10.1093/mnras/219.1.111}, \href
  {https://ui.adsabs.harvard.edu/abs/1986MNRAS.219..111B} {219, 111}

\bibitem[\protect\citeauthoryear{{Balona}, {Aerts}  \& {{\v{S}}tefl}}{{Balona}
  et~al.}{1999}]{Balona1999}
{Balona} L.~A.,  {Aerts} C.,   {{\v{S}}tefl} S.,  1999, \mn@doi [\mnras]
  {10.1046/j.1365-8711.1999.02386.x}, \href
  {https://ui.adsabs.harvard.edu/abs/1999MNRAS.305..519B} {305, 519}

\bibitem[\protect\citeauthoryear{{Blomme} et~al.,}{{Blomme}
  et~al.}{2011}]{Blomme2011}
{Blomme} R.,  et~al., 2011, \mn@doi [\aap] {10.1051/0004-6361/201116949}, \href
  {https://ui.adsabs.harvard.edu/abs/2011A&A...533A...4B} {533, A4}

\bibitem[\protect\citeauthoryear{{Bonanos}, {Castro}, {Macri}  \&
  {Kudritzki}}{{Bonanos} et~al.}{2011}]{Bonanos2011}
{Bonanos} A.~Z.,  {Castro} N.,  {Macri} L.~M.,   {Kudritzki} R.-P.,  2011,
  \mn@doi [\apjl] {10.1088/2041-8205/729/1/L9}, \href
  {https://ui.adsabs.harvard.edu/abs/2011ApJ...729L...9B} {729, L9}

\bibitem[\protect\citeauthoryear{{Bowman}}{{Bowman}}{2017}]{Bowman2017}
{Bowman} D.~M.,  2017, {Amplitude Modulation of Pulsation Modes in Delta Scuti
  Stars}.
Springer International Publishing, \mn@doi{10.1007/978-3-319-66649-5}

\bibitem[\protect\citeauthoryear{{Bowman}}{{Bowman}}{2020}]{Bowman2020b}
{Bowman} D.~M.,  2020, \mn@doi [Frontiers in Astronomy and Space Sciences]
  {10.3389/fspas.2020.578584}, \href
  {https://ui.adsabs.harvard.edu/abs/2020FrASS...7...70B} {7, 70}

\bibitem[\protect\citeauthoryear{{Bowman} et~al.,}{{Bowman}
  et~al.}{2019a}]{Bowman2019b}
{Bowman} D.~M.,  et~al., 2019a, \mn@doi [Nature Astronomy]
  {10.1038/s41550-019-0768-1}, \href
  {https://ui.adsabs.harvard.edu/abs/2019NatAs...3..760B} {3, 760}

\bibitem[\protect\citeauthoryear{{Bowman} et~al.,}{{Bowman}
  et~al.}{2019b}]{Bowman2019a}
{Bowman} D.~M.,  et~al., 2019b, \mn@doi [\aap] {10.1051/0004-6361/201833662},
  \href {https://ui.adsabs.harvard.edu/abs/2019A%26A...621A.135B} {621, A135}

\bibitem[\protect\citeauthoryear{{Bowman}, {Burssens}, {Sim{\'o}n-D{\'\i}az},
  {Edelmann}, {Rogers}, {Horst}, {R{\"o}pke}  \& {Aerts}}{{Bowman}
  et~al.}{2020}]{Bowman2020a}
{Bowman} D.~M.,  {Burssens} S.,  {Sim{\'o}n-D{\'\i}az} S.,  {Edelmann}
  P.~V.~F.,  {Rogers} T.~M.,  {Horst} L.,  {R{\"o}pke} F.~K.,   {Aerts} C.,
  2020, \mn@doi [\aap] {10.1051/0004-6361/202038224}, \href
  {https://ui.adsabs.harvard.edu/abs/2020A&A...640A..36B} {640, A36}

\bibitem[\protect\citeauthoryear{{Breger} et~al.,}{{Breger}
  et~al.}{1993}]{Breger1993b}
{Breger} M.,  et~al., 1993, \aap, \href
  {http://adsabs.harvard.edu/abs/1993A%26A...271..482B} {271, 482}

\bibitem[\protect\citeauthoryear{{Briquet} \& {Aerts}}{{Briquet} \&
  {Aerts}}{2003}]{Briquet2003}
{Briquet} M.,  {Aerts} C.,  2003, \mn@doi [\aap] {10.1051/0004-6361:20021683},
  \href {https://ui.adsabs.harvard.edu/abs/2003A&A...398..687B} {398, 687}

\bibitem[\protect\citeauthoryear{{Briquet}, {Morel}, {Thoul}, {Scuflaire},
  {Miglio}, {Montalb{\'a}n}, {Dupret}  \& {Aerts}}{{Briquet}
  et~al.}{2007}]{Briquet2007}
{Briquet} M.,  {Morel} T.,  {Thoul} A.,  {Scuflaire} R.,  {Miglio} A.,
  {Montalb{\'a}n} J.,  {Dupret} M.~A.,   {Aerts} C.,  2007, \mn@doi [\mnras]
  {10.1111/j.1365-2966.2007.12142.x}, \href
  {https://ui.adsabs.harvard.edu/#abs/2007MNRAS.381.1482B} {381, 1482}

\bibitem[\protect\citeauthoryear{{Briquet} et~al.,}{{Briquet}
  et~al.}{2011}]{Briquet2011}
{Briquet} M.,  et~al., 2011, \mn@doi [\aap] {10.1051/0004-6361/201015690},
  \href {https://ui.adsabs.harvard.edu/abs/2011A&A...527A.112B} {527, A112}

\bibitem[\protect\citeauthoryear{{Brott} et~al.,}{{Brott}
  et~al.}{2011}]{Brott2011a}
{Brott} I.,  et~al., 2011, \mn@doi [\aap] {10.1051/0004-6361/201016113}, \href
  {https://ui.adsabs.harvard.edu/abs/2011A&A...530A.115B} {530, A115}

\bibitem[\protect\citeauthoryear{{Burssens}, {Bowman}, {Aerts}, {Pedersen},
  {Moravveji}  \& {Buysschaert}}{{Burssens} et~al.}{2019}]{Burssens2019}
{Burssens} S.,  {Bowman} D.~M.,  {Aerts} C.,  {Pedersen} M.~G.,  {Moravveji}
  E.,   {Buysschaert} B.,  2019, \mn@doi [\mnras] {10.1093/mnras/stz2165},
  \href {https://ui.adsabs.harvard.edu/abs/2019MNRAS.489.1304B} {489, 1304}

\bibitem[\protect\citeauthoryear{{Burssens} et~al.,}{{Burssens}
  et~al.}{2020}]{Burssens2020}
{Burssens} S.,  et~al., 2020, \mn@doi [\aap] {10.1051/0004-6361/202037700},
  \href {https://ui.adsabs.harvard.edu/abs/2020A&A...639A..81B} {639, A81}

\bibitem[\protect\citeauthoryear{{Buysschaert} et~al.,}{{Buysschaert}
  et~al.}{2017}]{Buysschaert2017a}
{Buysschaert} B.,  et~al., 2017, \mn@doi [\aap] {10.1051/0004-6361/201630318},
  \href {http://adsabs.harvard.edu/abs/2017A%26A...602A..91B} {602, A91}

\bibitem[\protect\citeauthoryear{{Buysschaert}, {Aerts}, {Bowman}, {Johnston},
  {Van Reeth}, {Pedersen}, {Mathis}  \& {Neiner}}{{Buysschaert}
  et~al.}{2018}]{Buysschaert2018b}
{Buysschaert} B.,  {Aerts} C.,  {Bowman} D.~M.,  {Johnston} C.,  {Van Reeth}
  T.,  {Pedersen} M.~G.,  {Mathis} S.,   {Neiner} C.,  2018, \mn@doi [\aap]
  {10.1051/0004-6361/201832642}, \href
  {https://ui.adsabs.harvard.edu/#abs/2018A&A...616A.148B} {616, A148}

\bibitem[\protect\citeauthoryear{{Cantiello} et~al.,}{{Cantiello}
  et~al.}{2009}]{Cantiello2009}
{Cantiello} M.,  et~al., 2009, \mn@doi [\aap] {10.1051/0004-6361/200911643},
  \href {https://ui.adsabs.harvard.edu/abs/2009A&A...499..279C} {499, 279}

\bibitem[\protect\citeauthoryear{{Charbonneau}}{{Charbonneau}}{1995}]{Charbonneau1995}
{Charbonneau} P.,  1995, \mn@doi [\apjs] {10.1086/192242}, \href
  {https://ui.adsabs.harvard.edu/#abs/1995ApJS..101..309C} {101, 309}

\bibitem[\protect\citeauthoryear{{Choi}, {Dotter}, {Conroy}, {Cantiello},
  {Paxton}  \& {Johnson}}{{Choi} et~al.}{2016}]{Choi2016}
{Choi} J.,  {Dotter} A.,  {Conroy} C.,  {Cantiello} M.,  {Paxton} B.,
  {Johnson} B.~D.,  2016, \mn@doi [\apj] {10.3847/0004-637X/823/2/102}, \href
  {https://ui.adsabs.harvard.edu/#abs/2016ApJ...823..102C} {823, 102}

\bibitem[\protect\citeauthoryear{{Claret} \& {Bloemen}}{{Claret} \&
  {Bloemen}}{2011}]{Claret2011}
{Claret} A.,  {Bloemen} S.,  2011, \mn@doi [\aap]
  {10.1051/0004-6361/201116451}, \href
  {http://adsabs.harvard.edu/abs/2011A%26A...529A..75C} {529, A75}

\bibitem[\protect\citeauthoryear{{Daszy{\'n}ska-Daszkiewicz} \&
  {Walczak}}{{Daszy{\'n}ska-Daszkiewicz} \& {Walczak}}{2010}]{Daszynska2010}
{Daszy{\'n}ska-Daszkiewicz} J.,  {Walczak} P.,  2010, \mn@doi [\mnras]
  {10.1111/j.1365-2966.2009.16141.x}, \href
  {https://ui.adsabs.harvard.edu/abs/2010MNRAS.403..496D} {403, 496}

\bibitem[\protect\citeauthoryear{{Daszy{\'n}ska-Daszkiewicz}, {Szewczuk}  \&
  {Walczak}}{{Daszy{\'n}ska-Daszkiewicz} et~al.}{2013}]{Daszynska2013}
{Daszy{\'n}ska-Daszkiewicz} J.,  {Szewczuk} W.,   {Walczak} P.,  2013, \mn@doi
  [\mnras] {10.1093/mnras/stt418}, \href
  {https://ui.adsabs.harvard.edu/abs/2013MNRAS.431.3396D} {431, 3396}

\bibitem[\protect\citeauthoryear{{De Cat} \& {Aerts}}{{De Cat} \&
  {Aerts}}{2002}]{deCat2002}
{De Cat} P.,  {Aerts} C.,  2002, \mn@doi [\aap] {10.1051/0004-6361:20021068},
  \href {https://ui.adsabs.harvard.edu/abs/2002A&A...393..965D} {393, 965}

\bibitem[\protect\citeauthoryear{{De Cat}, {De Ridder}, {Hensberge}  \&
  {Ilijic}}{{De Cat} et~al.}{2004}]{deCat2004}
{De Cat} P.,  {De Ridder} J.,  {Hensberge} H.,   {Ilijic} S.,  2004, in
  {Hilditch} R.~W.,  {Hensberge} H.,   {Pavlovski} K.,  eds,  Astronomical
  Society of the Pacific Conference Series Vol. 318, Spectroscopically and
  Spatially Resolving the Components of the Close Binary Stars. pp 338--341

\bibitem[\protect\citeauthoryear{{Degroote} et~al.,}{{Degroote}
  et~al.}{2009}]{Degroote2009a}
{Degroote} P.,  et~al., 2009, \mn@doi [\aap] {10.1051/0004-6361/200911782},
  \href {http://adsabs.harvard.edu/abs/2009A%26A...506..111D} {506, 111}

\bibitem[\protect\citeauthoryear{{Degroote} et~al.,}{{Degroote}
  et~al.}{2010}]{Degroote2010}
{Degroote} P.,  et~al., 2010, \mn@doi [\nat] {10.1038/nature08864}, \href
  {https://ui.adsabs.harvard.edu/abs/2010Natur.464..259D} {464, 259}

\bibitem[\protect\citeauthoryear{{Donati}, {Semel}, {Carter}, {Rees}  \&
  {Collier Cameron}}{{Donati} et~al.}{1997}]{Donati1997}
{Donati} J.~F.,  {Semel} M.,  {Carter} B.~D.,  {Rees} D.~E.,   {Collier
  Cameron} A.,  1997, \mn@doi [\mnras] {10.1093/mnras/291.4.658}, \href
  {https://ui.adsabs.harvard.edu/abs/1997MNRAS.291..658D} {291, 658}

\bibitem[\protect\citeauthoryear{{Edelmann}, {Ratnasingam}, {Pedersen},
  {Bowman}, {Prat}  \& {Rogers}}{{Edelmann} et~al.}{2019}]{Edelmann2019}
{Edelmann} P.~V.~F.,  {Ratnasingam} R.~P.,  {Pedersen} M.~G.,  {Bowman} D.~M.,
  {Prat} V.,   {Rogers} T.~M.,  2019, \mn@doi [\apj]
  {10.3847/1538-4357/ab12df}, \href
  {https://ui.adsabs.harvard.edu/abs/2019ApJ...876....4E} {876, 4}

\bibitem[\protect\citeauthoryear{{Ekstr{\"o}m} et~al.,}{{Ekstr{\"o}m}
  et~al.}{2012}]{Ekstrom2012}
{Ekstr{\"o}m} S.,  et~al., 2012, \mn@doi [\aap] {10.1051/0004-6361/201117751},
  \href {https://ui.adsabs.harvard.edu/abs/2012A&A...537A.146E} {537, A146}

\bibitem[\protect\citeauthoryear{{Erdim} \& {H{\"u}daverdi}}{{Erdim} \&
  {H{\"u}daverdi}}{2019}]{Erdim2019}
{Erdim} M.~K.,  {H{\"u}daverdi} M.,  2019, in Turkish Physical Society 35th
  International Physics Congress (TPS35). p. 030023, \mn@doi{10.1063/1.5135421}

\bibitem[\protect\citeauthoryear{{Foreman-Mackey}, {Hogg}, {Lang}  \&
  {Goodman}}{{Foreman-Mackey} et~al.}{2013}]{ForemanMackey2013}
{Foreman-Mackey} D.,  {Hogg} D.~W.,  {Lang} D.,   {Goodman} J.,  2013, \mn@doi
  [\pasp] {10.1086/670067}, \href
  {http://adsabs.harvard.edu/abs/2013PASP..125..306F} {125, 306}

\bibitem[\protect\citeauthoryear{{Godart}, {Sim{\'o}n-D{\'{\i}}az}, {Herrero},
  {Dupret}, {Gr{\"o}tsch-Noels}, {Salmon}  \& {Ventura}}{{Godart}
  et~al.}{2017}]{Godart2017}
{Godart} M.,  {Sim{\'o}n-D{\'{\i}}az} S.,  {Herrero} A.,  {Dupret} M.~A.,
  {Gr{\"o}tsch-Noels} A.,  {Salmon} S.~J.~A.~J.,   {Ventura} P.,  2017, \mn@doi
  [\aap] {10.1051/0004-6361/201628856}, \href
  {http://adsabs.harvard.edu/abs/2017A%26A...597A..23G} {597, A23}

\bibitem[\protect\citeauthoryear{{Gray}}{{Gray}}{2005}]{Gray2005}
{Gray} D.~F.,  2005, {The Observation and Analysis of Stellar Photospheres}

\bibitem[\protect\citeauthoryear{{Handler}}{{Handler}}{2013}]{Handler2013}
{Handler} G.,  2013, {Asteroseismology}.
p.~207, \mn@doi{10.1007/978-94-007-5615-1_4}

\bibitem[\protect\citeauthoryear{Harris et~al.,}{Harris
  et~al.}{2020}]{Harris2020}
Harris C.~R.,  et~al., 2020, \mn@doi [Nature] {10.1038/s41586-020-2649-2}, 585,
  357

\bibitem[\protect\citeauthoryear{{Horst}, {Edelmann}, {Andr{\'a}ssy},
  {R{\"o}pke}, {Bowman}, {Aerts}  \& {Ratnasingam}}{{Horst}
  et~al.}{2020}]{Horst2020}
{Horst} L.,  {Edelmann} P.~V.~F.,  {Andr{\'a}ssy} R.,  {R{\"o}pke} F.~K.,
  {Bowman} D.~M.,  {Aerts} C.,   {Ratnasingam} R.~P.,  2020, \mn@doi [\aap]
  {10.1051/0004-6361/202037531}, \href
  {https://ui.adsabs.harvard.edu/abs/2020A&A...641A..18H} {641, A18}

\bibitem[\protect\citeauthoryear{{Howarth}, {Siebert}, {Hussain}  \&
  {Prinja}}{{Howarth} et~al.}{1997}]{Howarth1997}
{Howarth} I.~D.,  {Siebert} K.~W.,  {Hussain} G. A.~J.,   {Prinja} R.~K.,
  1997, \mn@doi [\mnras] {10.1093/mnras/284.2.265}, \href
  {https://ui.adsabs.harvard.edu/abs/1997MNRAS.284..265H} {284, 265}

\bibitem[\protect\citeauthoryear{{Howell} et~al.,}{{Howell}
  et~al.}{2014}]{Howell2014}
{Howell} S.~B.,  et~al., 2014, \mn@doi [\pasp] {10.1086/676406}, \href
  {https://ui.adsabs.harvard.edu/abs/2014PASP..126..398H} {126, 398}

\bibitem[\protect\citeauthoryear{Hunter}{Hunter}{2007}]{Hunter2007}
Hunter J.~D.,  2007, \mn@doi [Computing in Science \& Engineering]
  {10.1109/MCSE.2007.55}, 9, 90

\bibitem[\protect\citeauthoryear{{Ilijic}, {Hensberge}, {Pavlovski}  \&
  {Freyhammer}}{{Ilijic} et~al.}{2004}]{Ilijic2004}
{Ilijic} S.,  {Hensberge} H.,  {Pavlovski} K.,   {Freyhammer} L.~M.,  2004, in
  {Hilditch} R.~W.,  {Hensberge} H.,   {Pavlovski} K.,  eds,  Astronomical
  Society of the Pacific Conference Series Vol. 318, Spectroscopically and
  Spatially Resolving the Components of the Close Binary Stars. pp 111--113

\bibitem[\protect\citeauthoryear{{Johnston}, {Buysschaert}, {Tkachenko},
  {Aerts}  \& {Neiner}}{{Johnston} et~al.}{2017}]{Johnston2017}
{Johnston} C.,  {Buysschaert} B.,  {Tkachenko} A.,  {Aerts} C.,   {Neiner} C.,
  2017, \mn@doi [\mnras] {10.1093/mnrasl/slx060}, \href
  {https://ui.adsabs.harvard.edu/#abs/2017MNRAS.469L.118J} {469, L118}

\bibitem[\protect\citeauthoryear{{Johnston}, {Tkachenko}, {Aerts},
  {Molenberghs}, {Bowman}, {Pedersen}, {Buysschaert}  \&
  {P{\'a}pics}}{{Johnston} et~al.}{2019a}]{Johnston2019a}
{Johnston} C.,  {Tkachenko} A.,  {Aerts} C.,  {Molenberghs} G.,  {Bowman}
  D.~M.,  {Pedersen} M.~G.,  {Buysschaert} B.,   {P{\'a}pics} P.~I.,  2019a,
  \mn@doi [\mnras] {10.1093/mnras/sty2671}, \href
  {https://ui.adsabs.harvard.edu/abs/2019MNRAS.482.1231J} {482, 1231}

\bibitem[\protect\citeauthoryear{{Johnston}, {Pavlovski}  \&
  {Tkachenko}}{{Johnston} et~al.}{2019b}]{Johnston2019b}
{Johnston} C.,  {Pavlovski} K.,   {Tkachenko} A.,  2019b, \mn@doi [Astronomy
  and Astrophysics] {10.1051/0004-6361/201935235}, \href
  {https://ui.adsabs.harvard.edu/abs/2019A&A...628A..25J} {628, A25}

\bibitem[\protect\citeauthoryear{{Kiminki} \& {Kobulnicky}}{{Kiminki} \&
  {Kobulnicky}}{2012}]{Kiminki2012}
{Kiminki} D.~C.,  {Kobulnicky} H.~A.,  2012, \mn@doi [\apj]
  {10.1088/0004-637X/751/1/4}, \href
  {https://ui.adsabs.harvard.edu/abs/2012ApJ...751....4K} {751, 4}

\bibitem[\protect\citeauthoryear{{Kippenhahn}, {Weigert}  \&
  {Weiss}}{{Kippenhahn} et~al.}{2012}]{Kippenhahn2012}
{Kippenhahn} R.,  {Weigert} A.,   {Weiss} A.,  2012, {Stellar Structure and
  Evolution}.
Springer-Verlag Berlin Heidelberg, \mn@doi{10.1007/978-3-642-30304-3}

\bibitem[\protect\citeauthoryear{{Koumpia} \& {Bonanos}}{{Koumpia} \&
  {Bonanos}}{2012}]{Koumpia2012}
{Koumpia} E.,  {Bonanos} A.~Z.,  2012, \mn@doi [\aap]
  {10.1051/0004-6361/201219465}, \href
  {https://ui.adsabs.harvard.edu/abs/2012A&A...547A..30K} {547, A30}

\bibitem[\protect\citeauthoryear{{Krti{\v{c}}ka} \&
  {Feldmeier}}{{Krti{\v{c}}ka} \& {Feldmeier}}{2018}]{Krticka2018}
{Krti{\v{c}}ka} J.,  {Feldmeier} A.,  2018, \mn@doi [\aap]
  {10.1051/0004-6361/201731614}, \href
  {https://ui.adsabs.harvard.edu/abs/2018A&A...617A.121K} {617, A121}

\bibitem[\protect\citeauthoryear{{Kupka}, {Piskunov}, {Ryabchikova}, {Stempels}
   \& {Weiss}}{{Kupka} et~al.}{1999}]{Kupka1999}
{Kupka} F.,  {Piskunov} N.,  {Ryabchikova} T.~A.,  {Stempels} H.~C.,   {Weiss}
  W.~W.,  1999, \mn@doi [\aaps] {10.1051/aas:1999267}, \href
  {https://ui.adsabs.harvard.edu/abs/1999A&AS..138..119K} {138, 119}

\bibitem[\protect\citeauthoryear{{Labadie-Bartz} et~al.,}{{Labadie-Bartz}
  et~al.}{2020}]{LabadieBartz2020}
{Labadie-Bartz} J.,  et~al., 2020, \mn@doi [\aj] {10.3847/1538-3881/ab952c},
  \href {https://ui.adsabs.harvard.edu/abs/2020AJ....160...32L} {160, 32}

\bibitem[\protect\citeauthoryear{{Langer}}{{Langer}}{2012}]{Langer2012}
{Langer} N.,  2012, \mn@doi [\araa] {10.1146/annurev-astro-081811-125534},
  \href {https://ui.adsabs.harvard.edu/abs/2012ARA&A..50..107L} {50, 107}

\bibitem[\protect\citeauthoryear{{Lehmann}, {Tsymbal}, {Mkrtichian}  \&
  {Fraga}}{{Lehmann} et~al.}{2006}]{Lehmann2006}
{Lehmann} H.,  {Tsymbal} V.,  {Mkrtichian} D.~E.,   {Fraga} L.,  2006, \mn@doi
  [\aap] {10.1051/0004-6361:20065508}, \href
  {https://ui.adsabs.harvard.edu/abs/2006A&A...457.1033L} {457, 1033}

\bibitem[\protect\citeauthoryear{{Lohr}, {Clark}, {Najarro}, {Patrick},
  {Crowther}  \& {Evans}}{{Lohr} et~al.}{2018}]{Lohr2018}
{Lohr} M.~E.,  {Clark} J.~S.,  {Najarro} F.,  {Patrick} L.~R.,  {Crowther}
  P.~A.,   {Evans} C.~J.,  2018, \mn@doi [\aap] {10.1051/0004-6361/201832670},
  \href {https://ui.adsabs.harvard.edu/abs/2018A&A...617A..66L} {617, A66}

\bibitem[\protect\citeauthoryear{{Maeder}}{{Maeder}}{2009}]{Maeder2009}
{Maeder} A.,  2009, {Physics, Formation and Evolution of Rotating Stars}.
Springer-Verlag Berlin Heidelberg, \mn@doi{10.1007/978-3-540-76949-1}

\bibitem[\protect\citeauthoryear{{Mahy} et~al.,}{{Mahy}
  et~al.}{2020a}]{Mahy2020a}
{Mahy} L.,  et~al., 2020a, \mn@doi [\aap] {10.1051/0004-6361/201936151}, \href
  {https://ui.adsabs.harvard.edu/abs/2020A&A...634A.118M} {634, A118}

\bibitem[\protect\citeauthoryear{{Mahy} et~al.,}{{Mahy}
  et~al.}{2020b}]{Mahy2020b}
{Mahy} L.,  et~al., 2020b, \mn@doi [\aap] {10.1051/0004-6361/201936152}, \href
  {https://ui.adsabs.harvard.edu/abs/2020A&A...634A.119M} {634, A119}

\bibitem[\protect\citeauthoryear{{Mantegazza}, {Zerbi}  \&
  {Sacchi}}{{Mantegazza} et~al.}{2000}]{Mantegazza2000}
{Mantegazza} L.,  {Zerbi} F.~M.,   {Sacchi} A.,  2000, \aap, \href
  {https://ui.adsabs.harvard.edu/abs/2000A&A...354..112M} {354, 112}

\bibitem[\protect\citeauthoryear{{Martins}, {Schaerer}  \& {Hillier}}{{Martins}
  et~al.}{2005}]{Martins2005}
{Martins} F.,  {Schaerer} D.,   {Hillier} D.~J.,  2005, \mn@doi [\aap]
  {10.1051/0004-6361:20042386}, \href
  {https://ui.adsabs.harvard.edu/abs/2005A&A...436.1049M} {436, 1049}

\bibitem[\protect\citeauthoryear{{Mason}, {Gies}, {Hartkopf}, {Bagnuolo}, {ten
  Brummelaar}  \& {McAlister}}{{Mason} et~al.}{1998}]{Mason1998}
{Mason} B.~D.,  {Gies} D.~R.,  {Hartkopf} W.~I.,  {Bagnuolo} William~G. J.,
  {ten Brummelaar} T.,   {McAlister} H.~A.,  1998, \mn@doi [\aj]
  {10.1086/300234}, \href
  {https://ui.adsabs.harvard.edu/abs/1998AJ....115..821M} {115, 821}

\bibitem[\protect\citeauthoryear{{Maxted}}{{Maxted}}{2016}]{Maxted2016}
{Maxted} P.~F.~L.,  2016, \mn@doi [\aap] {10.1051/0004-6361/201628579}, \href
  {https://ui.adsabs.harvard.edu/abs/2016A&A...591A.111M} {591, A111}

\bibitem[\protect\citeauthoryear{{Maxted} et~al.,}{{Maxted}
  et~al.}{2020}]{Maxted2020}
{Maxted} P.~F.~L.,  et~al., 2020, \mn@doi [\mnras] {10.1093/mnras/staa1662},
  \href {https://ui.adsabs.harvard.edu/abs/2020MNRAS.498..332M} {498, 332}

\bibitem[\protect\citeauthoryear{{Mayer}, {Harmanec}  \& {Pavlovski}}{{Mayer}
  et~al.}{2013}]{Mayer2013}
{Mayer} P.,  {Harmanec} P.,   {Pavlovski} K.,  2013, \mn@doi [\aap]
  {10.1051/0004-6361/201220388}, \href
  {https://ui.adsabs.harvard.edu/abs/2013A&A...550A...2M} {550, A2}

\bibitem[\protect\citeauthoryear{{Mokiem}, {de Koter}, {Puls}, {Herrero},
  {Najarro}  \& {Villamariz}}{{Mokiem} et~al.}{2005}]{Mokiem2005}
{Mokiem} M.~R.,  {de Koter} A.,  {Puls} J.,  {Herrero} A.,  {Najarro} F.,
  {Villamariz} M.~R.,  2005, \mn@doi [\aap] {10.1051/0004-6361:20053522}, \href
  {https://ui.adsabs.harvard.edu/abs/2005A&A...441..711M} {441, 711}

\bibitem[\protect\citeauthoryear{{Moravveji}, {Townsend}, {Aerts}  \&
  {Mathis}}{{Moravveji} et~al.}{2016}]{Moravveji2016}
{Moravveji} E.,  {Townsend} R.~H.~D.,  {Aerts} C.,   {Mathis} S.,  2016,
  \mn@doi [\apj] {10.3847/0004-637X/823/2/130}, \href
  {https://ui.adsabs.harvard.edu/abs/2016ApJ...823..130M} {823, 130}

\bibitem[\protect\citeauthoryear{{Otero}}{{Otero}}{2007}]{Otero2007}
{Otero} S.~A.,  2007, Open European Journal on Variable Stars, \href
  {https://ui.adsabs.harvard.edu/abs/2007OEJV...72....1O} {0072, 1}

\bibitem[\protect\citeauthoryear{{Pamyatnykh}}{{Pamyatnykh}}{1999}]{Pamyatnykh1999}
{Pamyatnykh} A.~A.,  1999, \actaa, \href
  {https://ui.adsabs.harvard.edu/abs/1999AcA....49..119P} {49, 119}

\bibitem[\protect\citeauthoryear{{P{\'a}pics} et~al.,}{{P{\'a}pics}
  et~al.}{2012}]{Papics2012}
{P{\'a}pics} P.~I.,  et~al., 2012, \mn@doi [\aap]
  {10.1051/0004-6361/201218809}, \href
  {https://ui.adsabs.harvard.edu/abs/2012A&A...542A..55P} {542, A55}

\bibitem[\protect\citeauthoryear{{Pavlovski} \& {Hensberge}}{{Pavlovski} \&
  {Hensberge}}{2005}]{Pavlovski2005}
{Pavlovski} K.,  {Hensberge} H.,  2005, \mn@doi [\aap]
  {10.1051/0004-6361:20052804}, \href
  {https://ui.adsabs.harvard.edu/\#abs/2005A&A...439..309P} {439, 309}

\bibitem[\protect\citeauthoryear{{Paxton} et~al.,}{{Paxton}
  et~al.}{2015}]{Paxton2015}
{Paxton} B.,  et~al., 2015, \mn@doi [\apjs] {10.1088/0067-0049/220/1/15}, \href
  {https://ui.adsabs.harvard.edu/abs/2015ApJS..220...15P} {220, 15}

\bibitem[\protect\citeauthoryear{{Paxton} et~al.,}{{Paxton}
  et~al.}{2018}]{Paxton2018}
{Paxton} B.,  et~al., 2018, \mn@doi [\apjs] {10.3847/1538-4365/aaa5a8}, \href
  {https://ui.adsabs.harvard.edu/abs/2018ApJS..234...34P} {234, 34}

\bibitem[\protect\citeauthoryear{{Pedersen}, {Aerts}, {P{\'a}pics}  \&
  {Rogers}}{{Pedersen} et~al.}{2018}]{Pedersen2018}
{Pedersen} M.~G.,  {Aerts} C.,  {P{\'a}pics} P.~I.,   {Rogers} T.~M.,  2018,
  \mn@doi [\aap] {10.1051/0004-6361/201732317}, \href
  {https://ui.adsabs.harvard.edu/abs/2018A%26A...614A.128P} {614, A128}

\bibitem[\protect\citeauthoryear{{Pope} et~al.,}{{Pope}
  et~al.}{2019}]{Pope2019}
{Pope} B. J.~S.,  et~al., 2019, \mn@doi [\apjs] {10.3847/1538-4365/ab3d29},
  \href {https://ui.adsabs.harvard.edu/abs/2019ApJS..245....8P} {245, 8}

\bibitem[\protect\citeauthoryear{{Prsa}, {Matijevic}, {Latkovic}, {Vilardell}
  \& {Wils}}{{Prsa} et~al.}{2011}]{Prsa2011}
{Prsa} A.,  {Matijevic} G.,  {Latkovic} O.,  {Vilardell} F.,   {Wils} P.,
  2011, {PHOEBE: PHysics Of Eclipsing BinariEs}, Astrophysics Source Code
  Library (\mn@eprint {ascl} {1106.002})

\bibitem[\protect\citeauthoryear{{Pr{\v{s}}a} et~al.,}{{Pr{\v{s}}a}
  et~al.}{2016}]{Prsa2016}
{Pr{\v{s}}a} A.,  et~al., 2016, \mn@doi [\apjs] {10.3847/1538-4365/227/2/29},
  \href {https://ui.adsabs.harvard.edu/abs/2016ApJS..227...29P} {227, 29}

\bibitem[\protect\citeauthoryear{{Puls}, {Urbaneja}, {Venero}, {Repolust},
  {Springmann}, {Jokuthy}  \& {Mokiem}}{{Puls} et~al.}{2005}]{Puls2005}
{Puls} J.,  {Urbaneja} M.~A.,  {Venero} R.,  {Repolust} T.,  {Springmann} U.,
  {Jokuthy} A.,   {Mokiem} M.~R.,  2005, \mn@doi [\aap]
  {10.1051/0004-6361:20042365}, \href
  {https://ui.adsabs.harvard.edu/abs/2005A&A...435..669P} {435, 669}

\bibitem[\protect\citeauthoryear{{Puls}, {Vink}  \& {Najarro}}{{Puls}
  et~al.}{2008}]{Puls2008}
{Puls} J.,  {Vink} J.~S.,   {Najarro} F.,  2008, \mn@doi [\aapr]
  {10.1007/s00159-008-0015-8}, \href
  {https://ui.adsabs.harvard.edu/abs/2008A&ARv..16..209P} {16, 209}

\bibitem[\protect\citeauthoryear{{Raskin} et~al.,}{{Raskin}
  et~al.}{2011}]{Raskin2011}
{Raskin} G.,  et~al., 2011, \mn@doi [\aap] {10.1051/0004-6361/201015435}, \href
  {https://ui.adsabs.harvard.edu/#abs/2011A&A...526A..69R} {526, A69}

\bibitem[\protect\citeauthoryear{{Ratnasingam}, {Edelmann}  \&
  {Rogers}}{{Ratnasingam} et~al.}{2020}]{Ratnasingam2020}
{Ratnasingam} R.~P.,  {Edelmann} P.~V.~F.,   {Rogers} T.~M.,  2020, \mn@doi
  [\mnras] {10.1093/mnras/staa2296}, \href
  {https://ui.adsabs.harvard.edu/abs/2020MNRAS.497.4231R} {497, 4231}

\bibitem[\protect\citeauthoryear{{Ricker} et~al.,}{{Ricker}
  et~al.}{2015}]{Ricker2015}
{Ricker} G.~R.,  et~al., 2015, \mn@doi [Journal of Astronomical Telescopes,
  Instruments, and Systems] {10.1117/1.JATIS.1.1.014003}, \href
  {https://ui.adsabs.harvard.edu/#abs/2015JATIS...1a4003R} {1, 014003}

\bibitem[\protect\citeauthoryear{{Rivinius}, {Baade}  \&
  {{\v{S}}tefl}}{{Rivinius} et~al.}{2003}]{Rivinius2003}
{Rivinius} T.,  {Baade} D.,   {{\v{S}}tefl} S.,  2003, \mn@doi [\aap]
  {10.1051/0004-6361:20031285}, \href
  {https://ui.adsabs.harvard.edu/abs/2003A&A...411..229R} {411, 229}

\bibitem[\protect\citeauthoryear{{Rogers} \& {McElwaine}}{{Rogers} \&
  {McElwaine}}{2017}]{Rogers2017}
{Rogers} T.~M.,  {McElwaine} J.~N.,  2017, \mn@doi [\apjl]
  {10.3847/2041-8213/aa8d13}, \href
  {https://ui.adsabs.harvard.edu/abs/2017ApJ...848L...1R} {848, L1}

\bibitem[\protect\citeauthoryear{{Sana} et~al.,}{{Sana}
  et~al.}{2012}]{Sana2012}
{Sana} H.,  et~al., 2012, \mn@doi [Science] {10.1126/science.1223344}, \href
  {https://ui.adsabs.harvard.edu/abs/2012Sci...337..444S} {337, 444}

\bibitem[\protect\citeauthoryear{{Sana} et~al.,}{{Sana}
  et~al.}{2013}]{Sana2013}
{Sana} H.,  et~al., 2013, \mn@doi [\aap] {10.1051/0004-6361/201219621}, \href
  {https://ui.adsabs.harvard.edu/abs/2013A&A...550A.107S} {550, A107}

\bibitem[\protect\citeauthoryear{{Sana} et~al.,}{{Sana}
  et~al.}{2014}]{Sana2014}
{Sana} H.,  et~al., 2014, \mn@doi [\apjs] {10.1088/0067-0049/215/1/15}, \href
  {https://ui.adsabs.harvard.edu/abs/2014ApJS..215...15S} {215, 15}

\bibitem[\protect\citeauthoryear{{Sander}, {Vink}  \& {Hamann}}{{Sander}
  et~al.}{2019}]{Sander2019}
{Sander} A. A.~C.,  {Vink} J.~S.,   {Hamann} W.~R.,  2019, \mn@doi [\mnras]
  {10.1093/mnras/stz3064}, \href
  {https://ui.adsabs.harvard.edu/abs/2019MNRAS.tmp.2641S} {p.~2641}

\bibitem[\protect\citeauthoryear{{Schmid} \& {Aerts}}{{Schmid} \&
  {Aerts}}{2016}]{Schmid2016}
{Schmid} V.~S.,  {Aerts} C.,  2016, \mn@doi [\aap]
  {10.1051/0004-6361/201628617}, \href
  {https://ui.adsabs.harvard.edu/#abs/2016A&A...592A.116S} {592, A116}

\bibitem[\protect\citeauthoryear{{Schrijvers}, {Telting}, {Aerts}, {Ruymaekers}
   \& {Henrichs}}{{Schrijvers} et~al.}{1997}]{Schrijvers1997}
{Schrijvers} C.,  {Telting} J.~H.,  {Aerts} C.,  {Ruymaekers} E.,   {Henrichs}
  H.~F.,  1997, \mn@doi [\aaps] {10.1051/aas:1997321}, \href
  {https://ui.adsabs.harvard.edu/abs/1997A&AS..121..343S} {121, 343}

\bibitem[\protect\citeauthoryear{{Schrijvers}, {Telting}  \&
  {Aerts}}{{Schrijvers} et~al.}{2004}]{Schrijvers2004}
{Schrijvers} C.,  {Telting} J.~H.,   {Aerts} C.,  2004, \mn@doi [\aap]
  {10.1051/0004-6361:20031731}, \href
  {https://ui.adsabs.harvard.edu/abs/2004A&A...416.1069S} {416, 1069}

\bibitem[\protect\citeauthoryear{{Sim{\'o}n-D{\'\i}az} \&
  {Herrero}}{{Sim{\'o}n-D{\'\i}az} \& {Herrero}}{2014}]{SimonDiaz2014}
{Sim{\'o}n-D{\'\i}az} S.,  {Herrero} A.,  2014, \mn@doi [\aap]
  {10.1051/0004-6361/201322758}, \href
  {https://ui.adsabs.harvard.edu/abs/2014A&A...562A.135S} {562, A135}

\bibitem[\protect\citeauthoryear{{Sim{\'o}n-D{\'\i}az}, {Godart}, {Castro},
  {Herrero}, {Aerts}, {Puls}, {Telting}  \&
  {Grassitelli}}{{Sim{\'o}n-D{\'\i}az} et~al.}{2017}]{SimonDiaz2017}
{Sim{\'o}n-D{\'\i}az} S.,  {Godart} M.,  {Castro} N.,  {Herrero} A.,  {Aerts}
  C.,  {Puls} J.,  {Telting} J.,   {Grassitelli} L.,  2017, \mn@doi [\aap]
  {10.1051/0004-6361/201628541}, \href
  {https://ui.adsabs.harvard.edu/\#abs/2017A&A...597A..22S} {597, A22}

\bibitem[\protect\citeauthoryear{{Sim{\'o}n-D{\'\i}az}, {Aerts}, {Urbaneja},
  {Camacho}, {Antoci}, {Fredslund Andersen}, {Grundahl}  \&
  {Pall{\'e}}}{{Sim{\'o}n-D{\'\i}az} et~al.}{2018}]{SimonDiaz2018}
{Sim{\'o}n-D{\'\i}az} S.,  {Aerts} C.,  {Urbaneja} M.~A.,  {Camacho} I.,
  {Antoci} V.,  {Fredslund Andersen} M.,  {Grundahl} F.,   {Pall{\'e}} P.~L.,
  2018, \mn@doi [\aap] {10.1051/0004-6361/201732160}, \href
  {https://ui.adsabs.harvard.edu/abs/2018A&A...612A..40S} {612, A40}

\bibitem[\protect\citeauthoryear{{Southworth}}{{Southworth}}{2013}]{Southworth2013}
{Southworth} J.,  2013, \mn@doi [\aap] {10.1051/0004-6361/201322195}, \href
  {https://ui.adsabs.harvard.edu/abs/2013A&A...557A.119S} {557, A119}

\bibitem[\protect\citeauthoryear{{Sundqvist}, {Puls}, {Feldmeier}  \&
  {Owocki}}{{Sundqvist} et~al.}{2011}]{Sundqvist2011}
{Sundqvist} J.~O.,  {Puls} J.,  {Feldmeier} A.,   {Owocki} S.~P.,  2011,
  \mn@doi [\aap] {10.1051/0004-6361/201015771}, \href
  {https://ui.adsabs.harvard.edu/abs/2011A&A...528A..64S} {528, A64}

\bibitem[\protect\citeauthoryear{{Szewczuk} \&
  {Daszy{\'n}ska-Daszkiewicz}}{{Szewczuk} \&
  {Daszy{\'n}ska-Daszkiewicz}}{2017}]{Szewczuk2017}
{Szewczuk} W.,  {Daszy{\'n}ska-Daszkiewicz} J.,  2017, \mn@doi [\mnras]
  {10.1093/mnras/stx738}, \href
  {https://ui.adsabs.harvard.edu/abs/2017MNRAS.469...13S} {469, 13}

\bibitem[\protect\citeauthoryear{{Tkachenko}, {Van Reeth}, {Tsymbal}, {Aerts},
  {Kochukhov}  \& {Debosscher}}{{Tkachenko} et~al.}{2013}]{Tkachenko2013}
{Tkachenko} A.,  {Van Reeth} T.,  {Tsymbal} V.,  {Aerts} C.,  {Kochukhov} O.,
  {Debosscher} J.,  2013, \mn@doi [\aap] {10.1051/0004-6361/201322532}, \href
  {https://ui.adsabs.harvard.edu/abs/2013A&A...560A..37T} {560, A37}

\bibitem[\protect\citeauthoryear{{Tkachenko} et~al.,}{{Tkachenko}
  et~al.}{2020}]{Tkachenko2020}
{Tkachenko} A.,  et~al., 2020, \mn@doi [\aap] {10.1051/0004-6361/202037452},
  \href {https://ui.adsabs.harvard.edu/abs/2020A&A...637A..60T} {637, A60}

\bibitem[\protect\citeauthoryear{{Torres}, {Andersen}  \&
  {Gim{\'e}nez}}{{Torres} et~al.}{2010}]{Torres2010}
{Torres} G.,  {Andersen} J.,   {Gim{\'e}nez} A.,  2010, \mn@doi [ARA\&A]
  {10.1007/s00159-009-0025-1}, \href
  {https://ui.adsabs.harvard.edu/#abs/2010A&ARv..18...67T} {18, 67}

\bibitem[\protect\citeauthoryear{{Uytterhoeven}, {Telting}, {Aerts}  \&
  {Willems}}{{Uytterhoeven} et~al.}{2004}]{Uytterhoeven2004}
{Uytterhoeven} K.,  {Telting} J.~H.,  {Aerts} C.,   {Willems} B.,  2004,
  \mn@doi [\aap] {10.1051/0004-6361:20041224}, \href
  {https://ui.adsabs.harvard.edu/abs/2004A&A...427..593U} {427, 593}

\bibitem[\protect\citeauthoryear{{Uytterhoeven}, {Briquet}, {Aerts}, {Telting},
  {Harmanec}, {Lefever}  \& {Cuypers}}{{Uytterhoeven}
  et~al.}{2005}]{Uytterhoeven2005}
{Uytterhoeven} K.,  {Briquet} M.,  {Aerts} C.,  {Telting} J.~H.,  {Harmanec}
  P.,  {Lefever} K.,   {Cuypers} J.,  2005, \mn@doi [\aap]
  {10.1051/0004-6361:20041444}, \href
  {https://ui.adsabs.harvard.edu/abs/2005A&A...432..955U} {432, 955}

\bibitem[\protect\citeauthoryear{{Van Hamme} \& {Wilson}}{{Van Hamme} \&
  {Wilson}}{2007}]{VanHamme2007}
{Van Hamme} W.,  {Wilson} R.~E.,  2007, \mn@doi [\apj] {10.1086/517870}, \href
  {https://ui.adsabs.harvard.edu/abs/2007ApJ...661.1129V} {661, 1129}

\bibitem[\protect\citeauthoryear{{Vink}, {Muijres}, {Anthonisse}, {de Koter},
  {Gr{\"a}fener}  \& {Langer}}{{Vink} et~al.}{2011}]{Vink2011}
{Vink} J.~S.,  {Muijres} L.~E.,  {Anthonisse} B.,  {de Koter} A.,
  {Gr{\"a}fener} G.,   {Langer} N.,  2011, \mn@doi [\aap]
  {10.1051/0004-6361/201116614}, \href
  {https://ui.adsabs.harvard.edu/abs/2011A&A...531A.132V} {531, A132}

\bibitem[\protect\citeauthoryear{{Wade} et~al.,}{{Wade}
  et~al.}{2016}]{Wade2016}
{Wade} G.~A.,  et~al., 2016, \mn@doi [\mnras] {10.1093/mnras/stv2568}, \href
  {https://ui.adsabs.harvard.edu/abs/2016MNRAS.456....2W} {456, 2}

\bibitem[\protect\citeauthoryear{{Walczak} et~al.,}{{Walczak}
  et~al.}{2019}]{Walczak2019}
{Walczak} P.,  et~al., 2019, \mn@doi [\mnras] {10.1093/mnras/stz639}, \href
  {https://ui.adsabs.harvard.edu/abs/2019MNRAS.485.3544W} {485, 3544}

\bibitem[\protect\citeauthoryear{{White} et~al.,}{{White}
  et~al.}{2017}]{White2017b}
{White} T.~R.,  et~al., 2017, \mn@doi [\mnras] {10.1093/mnras/stx1050}, \href
  {https://ui.adsabs.harvard.edu/abs/2017MNRAS.471.2882W} {471, 2882}

\bibitem[\protect\citeauthoryear{{Wu} \& {Li}}{{Wu} \& {Li}}{2019}]{Wu2019}
{Wu} T.,  {Li} Y.,  2019, \mn@doi [\apj] {10.3847/1538-4357/ab2ad8}, \href
  {https://ui.adsabs.harvard.edu/abs/2019ApJ...881...86W} {881, 86}

\bibitem[\protect\citeauthoryear{{Zima}}{{Zima}}{2006}]{Zima2006a}
{Zima} W.,  2006, \mn@doi [\aap] {10.1051/0004-6361:20064876}, \href
  {https://ui.adsabs.harvard.edu/abs/2006A&A...455..227Z} {455, 227}

\bibitem[\protect\citeauthoryear{{Zima}}{{Zima}}{2008}]{Zima2008}
{Zima} W.,  2008, Communications in Asteroseismology, \href
  {https://ui.adsabs.harvard.edu/abs/2008CoAst.157..387Z} {157, 387}

\bibitem[\protect\citeauthoryear{{Zima} et~al.,}{{Zima}
  et~al.}{2006}]{Zima2006b}
{Zima} W.,  et~al., 2006, \mn@doi [\aap] {10.1051/0004-6361:20064877}, \href
  {https://ui.adsabs.harvard.edu/abs/2006A&A...455..235Z} {455, 235}

\bibitem[\protect\citeauthoryear{{de Mink}, {Langer}, {Izzard}, {Sana}  \& {de
  Koter}}{{de Mink} et~al.}{2013}]{deMink2013}
{de Mink} S.~E.,  {Langer} N.,  {Izzard} R.~G.,  {Sana} H.,   {de Koter} A.,
  2013, \mn@doi [\apj] {10.1088/0004-637X/764/2/166}, \href
  {https://ui.adsabs.harvard.edu/#abs/2013ApJ...764..166D} {764, 166}

\makeatother
\end{thebibliography}

\appendix
\section{Moment Method Periodograms}
\label{apdx:pergrams}
\begin{figure}
    \centering
    \includegraphics[width=0.9\columnwidth]{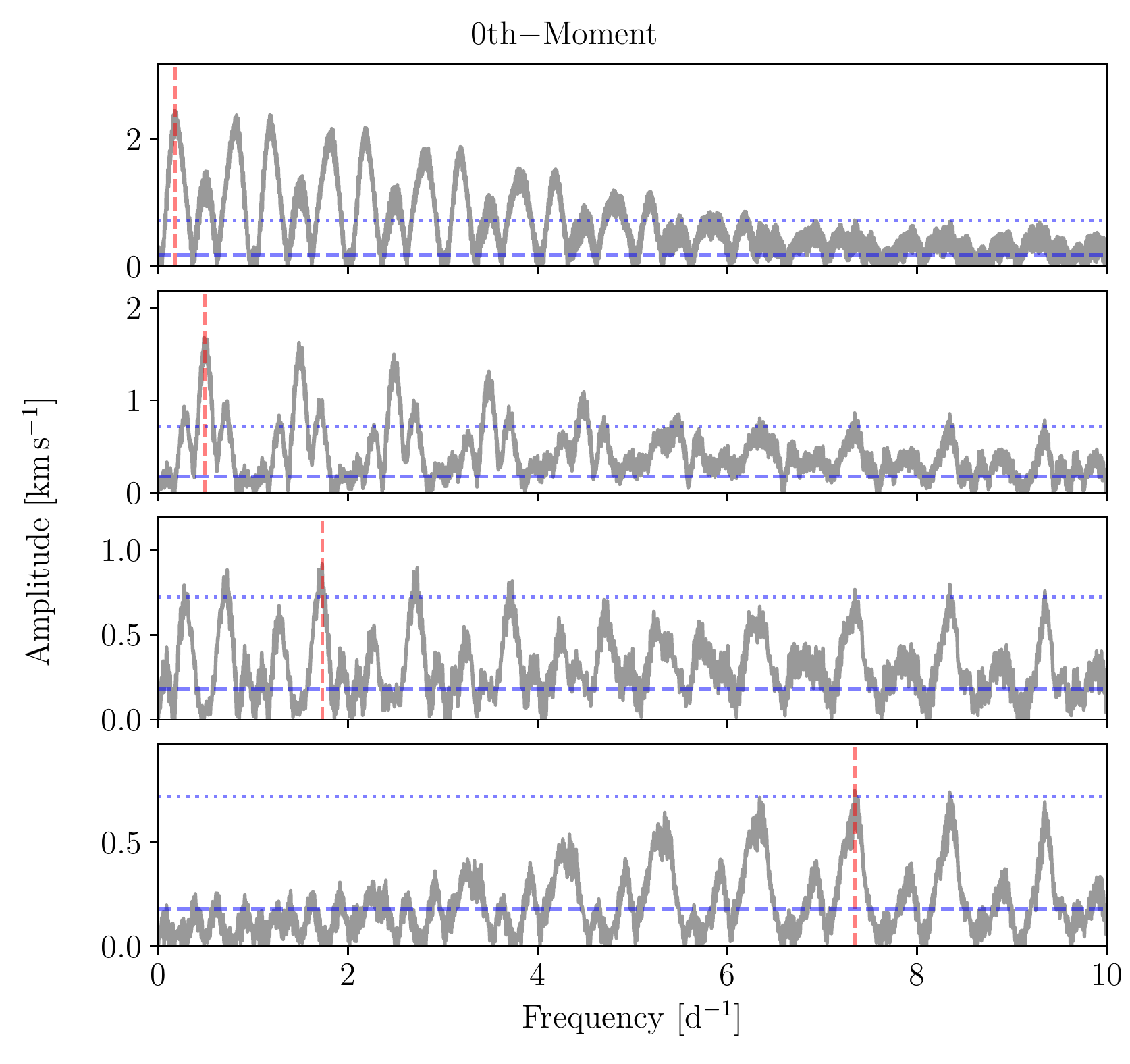}
    \caption{Periodograms of the 0th-moment of the LSD profiles. Extracted frequencies
             denoted by red vertical line. Dashed blue lines show noise level of
             pre-whitened data set, dotted blue lines denote four times the noise level.}
    \label{fig:0th_moment_periodogram}
\end{figure}

\begin{figure}
    \centering
    \includegraphics[width=0.9\columnwidth]{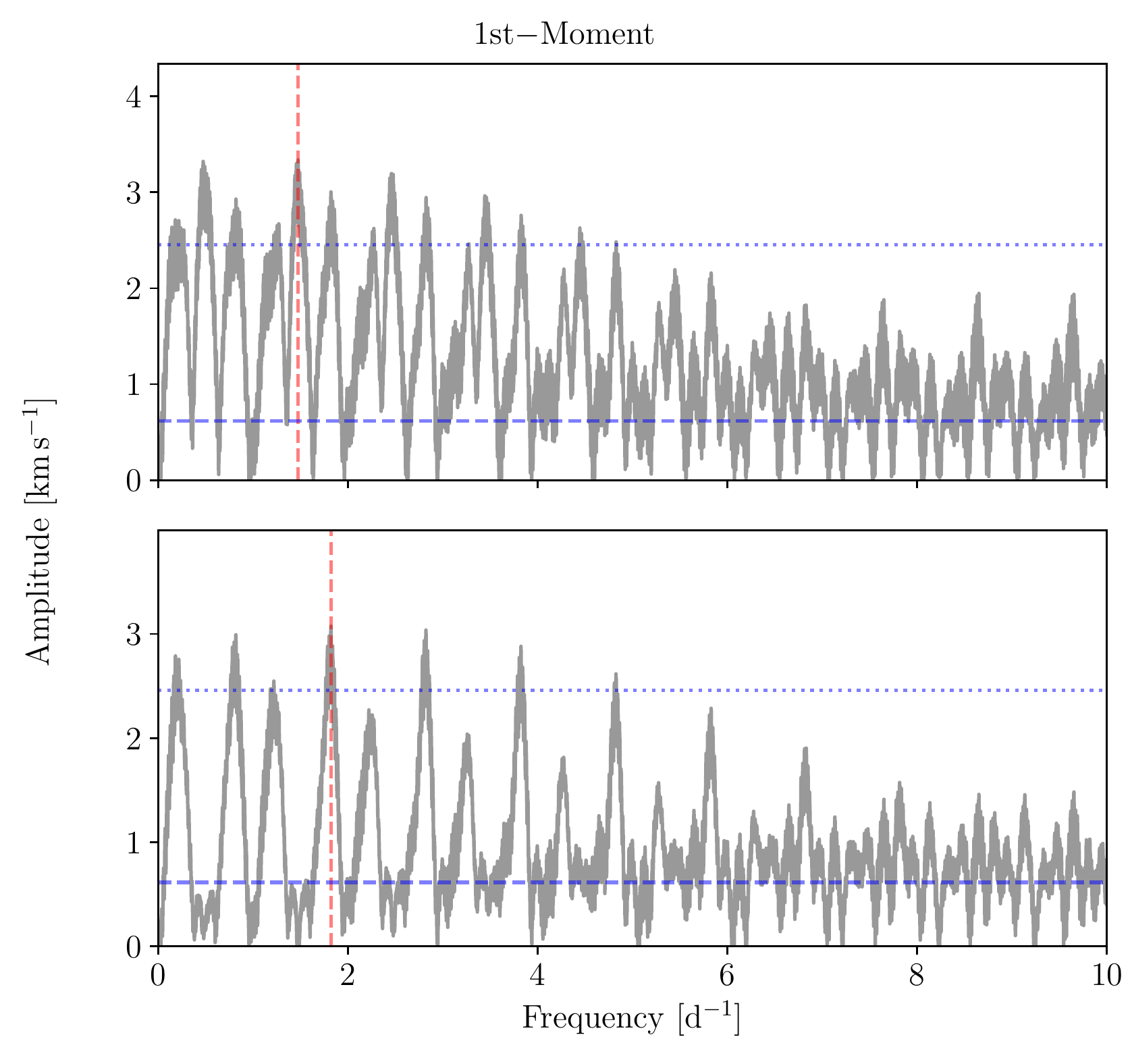}
    \caption{Same as Fig.~\ref{fig:0th_moment_periodogram} but for 1st moment.}
    \label{fig:1st_moment_periodogram}
\end{figure}

\begin{figure}
    \centering
    \includegraphics[width=0.9\columnwidth]{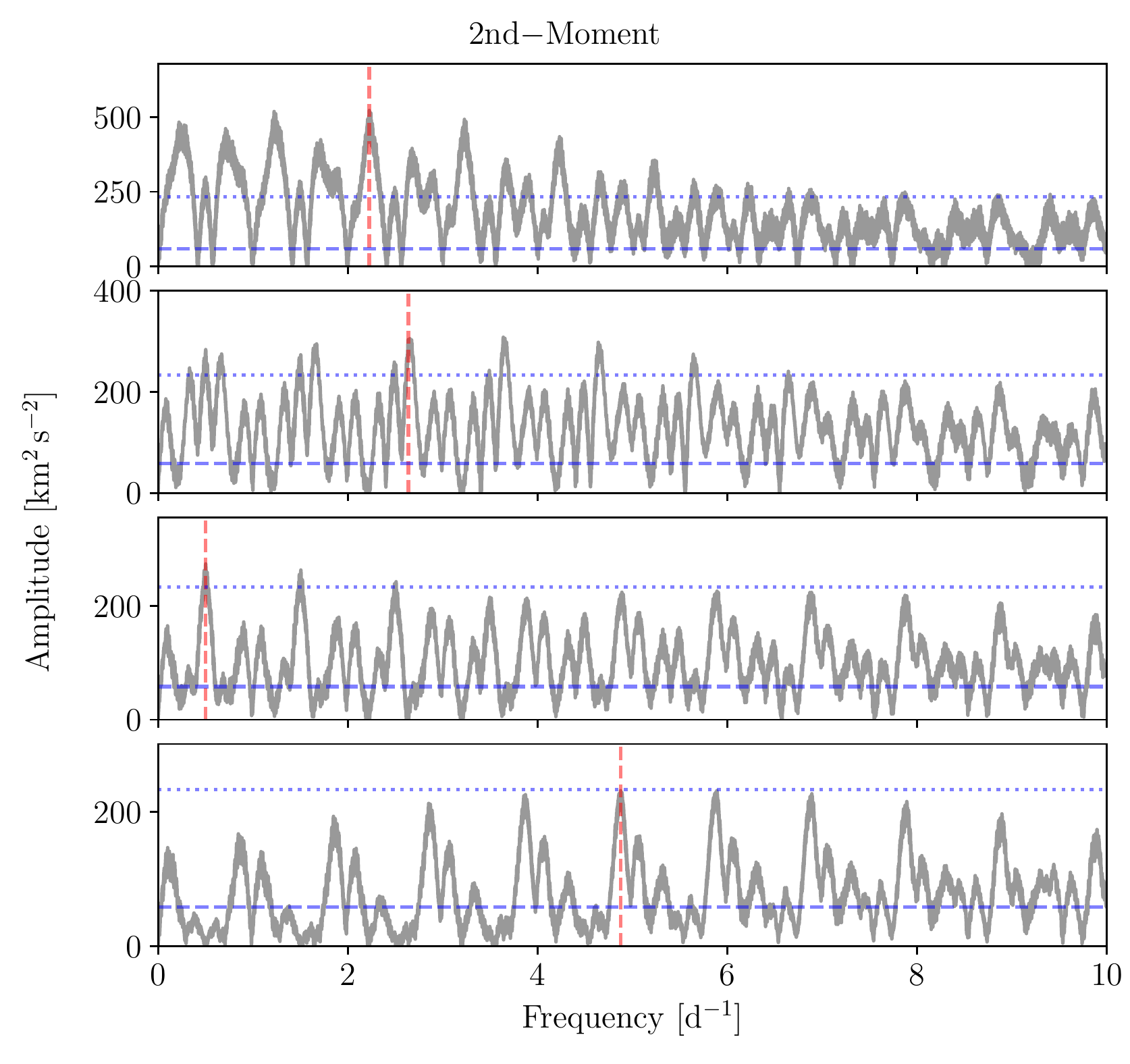}
    \caption{Same as Fig.~\ref{fig:0th_moment_periodogram} but for 2nd moment.}
    \label{fig:2nd_moment_periodogram}
\end{figure}

\begin{figure}
    \centering
    \includegraphics[width=0.9\columnwidth]{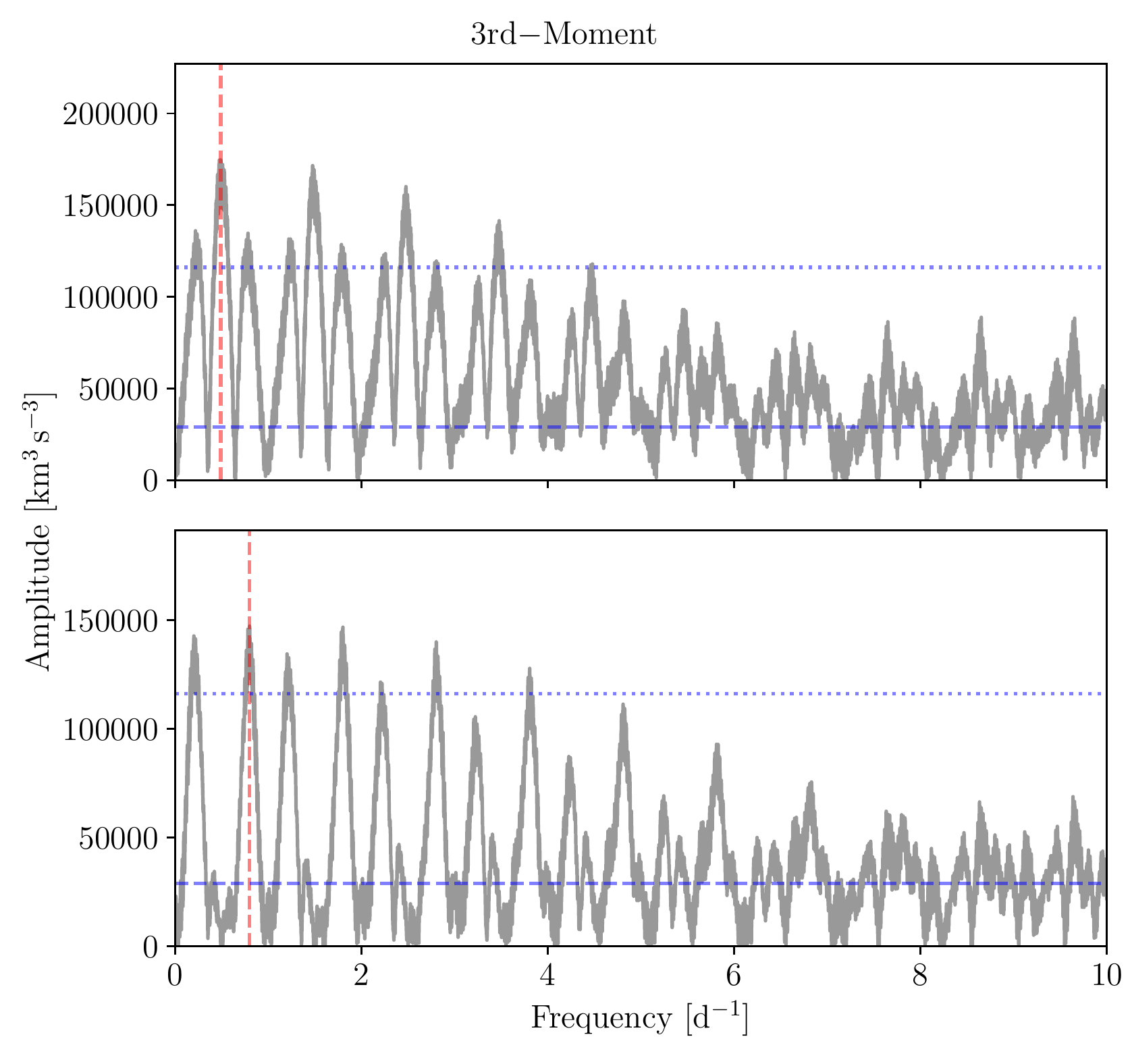}
    \caption{Same as Fig.~\ref{fig:0th_moment_periodogram} but for 3rd moment.}
    \label{fig:3rd_moment_periodogram}
\end{figure}

\bsp	
\label{lastpage}
\end{document}